\documentstyle[12pt,pageset,epsfig]{article}

\newcommand{\cc}[1]{\multicolumn{1}{c}{#1}}
\newcommand{\ccl}[1]{\multicolumn{1}{c|}{#1}}

\newcommand{\eps}{\epsilon}
 
\newcommand{\GeV}{{\,\rm GeV}}
\newcommand{\mubarn}{{\,\mu\rm barn}}

\newcommand{\RPP}{{\em RPP}}

\newcommand{\Qf}{{\rm Qf}}
\newcommand{\Pl}{{\rm Pl}}

\newcommand{\cP}{{\cal P}}

\begin{document}

\overfullrule 0pt
\tolerance 1600
\hbadness 1600
\vbadness 1600
\linepenalty=100

\rightline{BA--99--33}
\rightline{March 1999}

\vskip 0.75in
\begin{center}
{\bf SOPHIA}
\end{center}

\vskip .5cm

\begin{center}
{\bf Monte-Carlo simulations of photohadronic processes in astrophysics}
\end{center}

\vskip 0.5in

\begin{center}
A.~M\"ucke\footnote{University of Adelaide, Dept. Physics \& Math. Physics, 
Adelaide, SA 5005, Australia},
Ralph~Engel\footnote{Bartol Research Institute, University of Delaware, 
Newark, DE 19716, USA},
J.P.~Rachen\footnote{Pennsylvania State University, Dept. of Astronomy, 
University Park, PA 16802, USA}$^,$%
\footnote{Universiteit Utrecht, Sterrekundig Instituut, 
Princetonplein 5, 3584 CC Utrecht, The Netherlands},\\
R.J.~Protheroe$^{1}$
and
Todor~Stanev$^{2}$
\end{center}

\vskip 1in

\begin{center}
{\bf Abstract}
\end{center}

 \noindent A new Monte Carlo program for photohadronic 
 interactions of relativistic nucleons with an ambient photon radiation
 field is presented. The event generator is designed to fulfil
 typical astrophysical requirements, but can also be used for
radiation and background studies at high energy colliders such
as LEP2 and HERA, as well as for simulations of photon induced air showers. 
 We consider the full photopion production cross
 section from the pion production threshold up to high energies.
 It includes resonance excitation and decay, direct single pion
 production and  diffractive and non--diffractive multiparticle production.
 The cross section of each individual process is calculated by
 fitting experimental data, while the kinematics is determined by the
 underlying particle production process. We demonstrate that our model
 is capable of reproducing known accelerator data over a wide
 energy range.

\vskip.8cm
 

\noindent {\bf PACS:} \hspace*{.75cm} 13.85.Tp, 13.60.-r, 13.60.Hb
\hspace*{2.25cm} photoproduction cross section, Monte Carlo event generator,\\
\hspace*{2.25cm} multiparticle production

\vfill\eject
\section{Program Summary}

$$\vbox{
        \halign{#\hfil\quad &
                \vtop{\hsize 3.5in \strut # \strut}\cr
Title of program:               & SOPHIA 2.0\cr
\noalign{\vskip 10pt}
Catalog number:                 & \cr
\noalign{\vskip 10pt}
Program obtainable              & from A.~M\"ucke\cr
                                & e-mail: amuecke@physics.adelaide.edu.au\cr
\noalign{\vskip 10pt}
Computer on which the program   & DEC-Alpha and Intel-Pentium based\cr
has been thoroughly tested:     & workstations\cr
\noalign{\vskip 10pt}
Operating system:               & UNIX, Linux, Open-VMS\cr
\noalign{\vskip 10pt}
Programming language used:      & FORTRAN 77\cr
\noalign{\vskip 10pt}
Memory required to execute      & $<$1 megabyte \cr
\noalign{\vskip 10pt}
No. of bits in a word           & 64 \cr
 \noalign{\vskip 10pt}
Has the code been vectorized?   & no \cr
\noalign{\vskip 10pt}
Number of lines                 & 16500\cr
in distributed program:         & \cr
\noalign{\vskip 10pt}
Other programs used in SOPHIA   & {\sc Rndm}\cr
in modified form                & Processor independent random number\cr
                                & generator based on 
                                  Ref.~\cite{Marsaglia87}\cr
                                & {\sc Jetset 7.4}\cr
                                & Lund Monte Carlo for\cr
                                & jet fragmentation \cite{Sjostrand94a}\cr
                                & DECSIB \cr
                                & {\sc Sibyll} routine which decays\cr
                                & unstable particles \cite{Fletcher94}\cr
\noalign{\vskip 10pt}
Nature of physical problem:     & Simulation of minimum bias photo-\cr
                                & production for astrophysical applications\cr
}   
}$$ 

\clearpage

$$\vbox{
        \halign{#\hfil\quad &
                \vtop{\hsize 3.5in \strut # \strut}\cr

Method of solution:             & Monte Carlo simulation of individual events\cr
                                & for given nucleon and photon energies,\cr
                                & photon energies are sampled\cr
                                & from various distributions.\cr
\noalign{\vskip 10pt}
Restrictions on the complexity  & Incident ambient photon fields limited\cr
of the problem                  & to power law and blackbody spectra so far.\cr
                                & No tests were done for center-of-mass\cr
                                & energies $\sqrt s \ge 1000~$GeV.\cr
\noalign{\vskip 10pt}
Typical running time:           & 10000 events at a center-of-mass energy\cr
                                & of 1.5 GeV require a typical\cr
                                & CPU time of about $75$ seconds.\cr
}   
}$$ 

\clearpage

\section{Introduction}

 The cosmic ray spectrum extends to extremely high energies. Giant air
 showers have been observed with energy exceeding $\simeq 10^{11}$ GeV
 \cite{Takeda98a,Birdetal}. Energy losses due to interactions with ambient
 photons can become important, even dominant for such energetic nucleons,
 above the threshold for pion production. Photoproduction of hadrons is
 expected to cause a distortion of the ultra-high energy cosmic ray (CR)
 spectrum by interactions of the nucleons with the microwave background (the
 Greisen-Zatsepin-Kuzmin cutoff \cite{grei66,zats66}; see also \cite{prot96a}
 for additional references), but it may also be relevant to the observed high
 energy gamma ray emission from jets of Active Galactic Nuclei (AGN)
 (e.g. \cite{mann93,prot96b}) or Gamma-Ray Bursts (GRB)
 \cite{boet98}. Moreover, it is the major source process for the predicted
 fluxes of very high energy cosmic neutrinos (e.g.~\cite{prot96a,mann96,
 halz97,rach98,waxm99,Mannheim99a}).

 The photohadronic cross section at low interaction energies is dominated
 by the $\Delta$(1232) resonance. Since the low energy region of the cross
 section is emphasized in many astrophysical applications, the cross section
 and decay properties of the prominent $\Delta$-resonance have often been
 used as an approximation for photopion production, and the subsequent
 production of gamma rays and neutrinos \cite{stec73,gais95}. As discussed in
 \cite{Rach96,Mucke98a}, this approximation is only valid for a
 restricted number of cases, and does not describe sufficiently well the whole
 energy range of photohadronic interactions.  A more sophisticated
 photoproduction simulation code is needed to cover the center-of-mass energy
 range of about $\sqrt{s}\approx 1 - 10^3$ GeV, which is important in many
 astrophysical applications.

 In this paper we present the newly developed Monte-Carlo event generator
 SOPHIA (\underline{S}imulations \underline{O}f \underline{P}hoto
 \underline{H}adronic \underline{I}nteractions in \underline{A}strophysics),
 which we wrote as a tool for solving problems connected to photohadronic
 processes in astrophysical environments. The philosophy of the development
 of SOPHIA has been to implement  well established phenomenological models,
symmetries of hadronic interactions in a way that
 describes correctly  the available exclusive and inclusive photohadronic
 cross section data obtained at fixed target and collider experiments. 

 The paper is organized as follows. After introducing the kinematics of
 $N\gamma$-interactions (Sect.~3) we give a brief physical description of the
 relevant photohadronic interaction processes (Sect.~4). The implementation
 of these processes into the SOPHIA event generator, together with the method
 of cross section decomposition and parametrization, is described in
 Sect.~5. The structure of the program is outlined in Sec.~6, and a
 comparison of the model results with experimental data is provided in
 Sec.~7. The definitions of special functions and tables of parameters used in
 the cross section and final state parametrization, as well as a compilation
 of the important routines and functions used in the code, are given in the
 appendices. 

Unless noted otherwise, natural units ($\hbar = c = e = 1$) are used
throughout this paper, with GeV as the general unit. In this notation, cross
sections will be in GeV$^{-2}$. A general exception is Section 4, where
numerical parametrizations of cross sections are given in $\mu$barn. The
relevant conversion constant is $(\hbar c)^2 = 389.37966$ GeV$^{-2} \mu$barn.

\section{The physics of photohadronic interactions\label{sect:physics}}

\subsection{Kinematics of $N$-$\gamma$ collisions}

 There are three reference frames involved in the
 description of an astrophysical photohadronic interaction: (i) the
 astrophysical lab frame (LF), (ii) the rest frame of the nucleon%
\footnote{In the following the subscript $N$ is used if no distinction
between protons and neutrons is made. Interactions for both protons and
neutrons are implemented in SOPHIA.}
 (NRF), and (iii) the center-of-mass frame (CMF) of the interaction. For
example, in the
 LF, the initial state can be characterized by the nucleon energy $E_N$,
the photon energy $\epsilon$, and the interaction angle $\theta$
\begin{equation}
\cos \theta = ({\vec p}_N\cdot{\vec p}_\gamma)/\beta_N E_N \eps\ .
\label{costheta}
\end{equation}
where ${\vec p}_N$ and ${\vec p}_\gamma$ denote the nucleon and photon
momenta.  The Lorentz factor is $\gamma_N = E_N/m_N =
(1-\beta_N^2)^{-1/2}$ with $m_N$ being the
nucleon mass. The corresponding quantities in the NRF and CMF are marked with
a prime ($^\prime$) and an asterisk ($^*$), respectively.  Fixed target
accelerator experiments where a photon beam interacts with a proton target
are performed in the NRF. In astronomical applications we assume that the LF
can be chosen such that a the photon distribution function is isotropic. The
LF may therefore be different from the astronomical observer's frame if, for
example the emission region is moving relative to the observer such as in AGN
jets or GRB. The interaction rate of the nucleon in the LF is given by
\begin{equation}
R(E_N) = \frac{1}{8E_N^2\beta_N} \int_{\epsilon_{\rm th}}^{\infty}\!d\epsilon\,
\frac{n(\epsilon)}{\epsilon^2} \int_{s_{\rm th}}^{s_{\rm max}}\!ds\,(s-m_N^2)
\sigma_{N\gamma}(s)\ ,
\label{intrate}
\end{equation}
 where $\sigma_{N\gamma}$ is the total photohadronic cross section and 
\begin{equation}
s = m_N^2 + 2E_N\eps (1-\beta_N\cos\theta) = m_N^2 + 2m_N\eps' ,
\label{CMF}
\end{equation}
is the square of the center-of-mass energy. The lowest threshold energy
for photomeson production is $\sqrt{s_{\rm th}} = m_N + m_{\pi^0}$. The
remaining quantities in Eq.~(\ref{intrate}) are $\eps_{\rm th} = (s_{\rm
th}-m_N^2)/2 (E_N+p_N)$ and $s_{\rm max} = m_N^2+ 2E_N\epsilon(1+\beta_N)$.

The final state of the interaction is described by a number $N_c$ of
possible channels, each of which has $N_{f,c}$ particles in the final
state. At threshold, the phase space volume vanishes, which requires
kinematically for the partial cross section $\sigma_c\to 0$ for $s\to
s_{{\rm th},c} = [\sum_i m_i]^2$. Above threshold, each final state
channel has $3N_{f,c}-4$ degrees of freedom given by the 3-momentum
components $(p_i,\chi_i,\phi_i)$ of the particles, constrained by energy
and momentum conservation. Here $p_i$, $\chi_i$, and $\phi_i$ are the
particle momentum, and it's polar and azimuthal angles with respect to
the
initial nucleon momentum, respectively.  One of the $\phi_i$ angles can
be
chosen to be distributed isotropically since we consider only the
scattering of unpolarized photons and nucleons, all other variables are
determined by the interaction physics through the differential cross
sections.  A distinguished role in the final state is played by the
``leading-baryon'', which is considered to carry the baryonic quantum
numbers of the incoming nucleon. For this particle, the Lorentz
invariant
4-momentum transfer $t=(P_N- P_{\rm final})^2$ is often used as a final
state variable. At small $s$, many interaction channels can be reduced
to
2-particle final states, for which $d\sigma/dt$ gives a complete
description.

\subsection{Interaction processes\label{chap:intproc}}

Photon-proton interactions are dominated by resonance production at low
energies. The incoming baryon is excited to a baryonic resonance due to
the absorption of the photon. Such resonances have very short life times
and decay immediately into other hadrons. Consequently, the $N\gamma$ cross
section exhibits a strong energy dependence with clearly visible
resonance peaks.
Another process being important 
at low energy is the incoherent interaction
of photons with the virtual structure of the nucleon. This process is
called direct meson production. 
Eventually, at high interaction energies ($\sqrt{s} > 2\GeV$) 
the total interaction
cross section becomes approximately energy-independent, 
while the contributions
from resonances and the direct interaction channels decrease.
In this energy range, photon-hadron interactions are dominated by 
inelastic multiparticle production (also called  multipion production).

\subsubsection{Baryon resonance excitation and decay}

 The energy range from the photopion threshold energy $\sqrt{s}_{\rm th}
\approx
 1.08$ GeV for $\gamma N$-interactions up to
 $\sqrt{s}\approx 2$ GeV is dominated by the process of resonant absorption
 of a photon by the nucleon with the subsequent emission of particles,
 i.e. the excitation and decay of baryon resonances. 
The cross section for the production of a resonance with angular momentum
$J$ is given by the Breit-Wigner formula
\begin{equation}
\sigma_{\rm bw}(s;M,\Gamma,J) = \frac{s}{(s-m_N^2)^2}\, 
\frac{4 \pi b_\gamma (2J+1) s \Gamma^2}{(s-M^2)^2 + s \Gamma^2}\ ,
\label{breitwigner}
\end{equation}
where $M$ and $\Gamma$ are the nominal mass and the width of the resonance.
$b_\gamma$ is the branching ratio for photo-decay of the resonance, which is
identical to the probability of photoexcitation. The decay of baryon
resonances is generally dominated by hadronic channels. The exclusive cross
sections for the resonant contribution to a hadronic channel with branching
ratio $b_c$ can be written as
\begin{equation}
\sigma_c(s;M,\Gamma,J) = b_c \sigma_{\rm bw}(s; M,\Gamma,J),
\end{equation}
with $\sum_c b_c = 1 - b_\gamma \approx 1$. Most decay channels produce
 two-particle intermediate or final states, some of them again involving
 resonances. For the pion-nucleon decay channel, $N\pi$, the angular
 distribution of the final state is given by
\begin{equation}
\frac{d\sigma_{N\pi}}{d\cos\chi^*} \propto \sum_{\lambda = -J}^J
	\left| f_{\frac12,\lambda}^J
	d_{\lambda,\frac12}^J(\chi^*)\right|^2\ ,
\label{decay-distributions}
\end{equation}   
where $\chi^*$ denotes the scattering angle in the CMF
and $f_{\frac12,\lambda}^J$ are the $N\pi$-helicity amplitudes. The
functions $d_{\lambda,\frac12}^J(\chi^*)$ are commonly used angular 
distribution functions which are defined on the basis of spherical harmonics.
The
 $N\pi$ helicity amplitudes can be determined from the helicity amplitudes
 $A_{\frac12}$ and $A_{\frac32}$ for photoexcitation (see Ref.~\cite{Bransden73}
 for details), which are measured for many baryon
 resonances \cite{Caso98a}. The same expression applies to other final states
 involving a nucleon and an isospin-0 meson (e.g., $N\eta$). For decay
 channels with other spin parameters, however, the situation is more
 complex, and we assume for simplicity an isotropic decay of the
 resonance.

Baryon resonances are distinguished by their isospin into $N$-resonances
($I=\frac12$, as for the unexcited nucleon) and $\Delta$-resonances
($I=\frac32$).  The charge branching ratios $b_{\rm iso}$ of the resonance
decay follow from isospin symmetry.  For example, the branching ratios for
the decay into a two-particle final state involving a $N$- or
$\Delta$-baryon and an $I=1$ meson ($\pi$ or $\rho$) are given in
Table~\ref{tab:iso}. Here $\Delta I_3$ is the difference in the isospin
3-component of the baryon between initial and final state (the baryon charge
is $Q_B = I_3 + \frac12$).  In contrast to the strong decay channels, the
electromagnetic excitation of the resonance does not conserve isospin. Hence,
the resonance excitation strengths for $p\gamma$ and $n\gamma$ interactions
are not related to each other by isospin symmetry and have to be determined
experimentally.
\begin{table}[htb]
 \caption[]{\label{tab:iso} Isospin (charge) branching ratios for the decay
 of a resonance with isospin $I_{\rm res}$ and charge $I_3+\frac 12$ into a
 final state containing a baryon $B_f$ with isospin $I_{B_f}$ and charge $I_3 +
 \Delta I_3 + \frac12$, and a meson of isospin $1$ ($\pi$ or $\rho$). For
 example, the decay $N^+\to \Delta^{++}\pi^-$ corresponds to $I_{\rm res} =
 \frac12$, $I_{B_f} = \frac32$, $I_3 = \frac12$, and $\Delta I_3 = +1$, thus
 $I_3\,\Delta I_3 > 0$ and $b_{\rm iso} = \frac12$.}
\begin{center}
\renewcommand{\arraystretch}{1.2}
\begin{tabular}{|c|cc|ccc|}\hline
$b_{\rm iso} =$ & \multicolumn{2}{c|}{$I_{B_f} = 1/2$ ($B_f=N$)} &   
\multicolumn{3}{c|}{$I_{B_f} = 3/2$ ($B_f=\Delta$)}\\\hline 
	$I_{\rm res}$ & $\Delta I_3 \neq 0$ & $\Delta I_3=0$ & 
	$I_3\, \Delta I_3 < 0$ & $\Delta I_3 = 0$ & $I_3\,\Delta I_3 > 0$
	\\\hline  
	$1/2$ & $2/3$ & $1/3$ & $1/6$  & $1/3$  &  $1/2$ \\
	$3/2$ & $1/3$ & $2/3$ & $8/15$ & $1/15$ &  $2/5$ \\\hline
\end{tabular}
\end{center}
\end{table}

\subsubsection{Direct pion production}

Direct pion production can be considered as
 electromagnetic scattering by virtual charged mesons, which are the
 quantum--mechanical representation of the (color-neutral) strong force
 field around the baryon. The interacting virtual meson gains enough
 momentum to materialize.  Experimentally the direct production of charged
 pions is observed as a relatively  structureless background in the
 $N\pi^\pm$ and $\Delta\pi^\pm$ final states in photon-nucleon interactions. 

 In terms of Feynman graphs, this process is represented by the $t$-channel
exchange of a meson. Here, $t$ is the squared 4-momentum transfer from the
initial to the final state baryon.  The graph
 has a strong vertex at the baryon branch and an electromagnetic vertex for
 the photon interaction.  At the strong vertex, the baryon may be excited and
 change its isospin.  Isospin combination rules determine the iso-branching
 ratios in the same way as for resonance decay (Table~\ref{tab:iso} for
 $I_{\rm res} = \frac12$).  The presence of the electromagnetic vertex
 requires that the particle the photon couples to is charged. Thus direct
 processes with $\Delta I_3=0$ branches (e.g. $\gamma p \rightarrow \pi^0 p$)
 are strongly suppressed.

 The low energy structure of the direct cross section is not well
constrained. At high energies, Regge theory of the pion exchange implies 
that $\sigma_{\rm dir}(s) \propto
s^{-2}$ \cite{Collins77,Donnachie78a}.  
The angular distribution of the process is strongly forward peaked
and can be parametrized for small $|t|$ by
\begin{equation}
\frac{d\sigma_{\rm dir}}{dt} \propto \exp(b_{\rm dir} t)\ .
\label{t-dist}
\end{equation}  
with an experimentally determined slope of 
$b_{\rm dir}\approx12\GeV^{-2}$ \cite{Donnachie78a}.

The total cross section for a direct scattering process is roughly $\propto
m_t^{-2}$, where $m_t$ is the (nominal) mass of the exchanged virtual
 particle. Therefore, the direct production of pions is dominant, while the
 contributions from the exchange of heavier mesons are suppressed. The same
 applies to direct reactions which involve the exchange of a virtual baryon
 ($u$-channel exchange). However, with increasing energy, more and more
 channels add to the direct cross section, and this makes an explicit treatment
 difficult.

\subsubsection{High energy processes}

Phenomenologically, high energy interactions can be interpreted as reggeon
and pomeron exchange processes. Both the reggeon and the pomeron are
quasi-particles which correspond to sums of certain Feynman diagrams
in the Regge limit ($|t| \ll s$) \cite{Collins77}. 
The cross sections for reggeon and pomeron exchange have different but
universal energy 
dependences and account for all of the
total cross section \cite{Donnachie92b} at high energy. 
There are many different Regge theory-based cross section 
parametrizations possible. 
Here we use a recent cross section fit \cite{Caso98a} 
based on the Donnachie-Landshoff model \cite{Donnachie92b}
\begin{equation}
\label{sigma-reg} \sigma_{\rm reg} \propto 
\left(\frac{s-m_p^2}{s_0}\right)^{-0.34}
\hspace*{2cm}
\label{sigma-pom} \sigma_{\rm pom} \propto 
\left(\frac{s-m_p^2}{s_0}\right)^{0.095}\ ,
\label{DL-model}
\end{equation}
with the reference scale $s_0 = 1$ GeV$^2$.

Concerning high energy processes, it is convenient to
distinguish between diffractive and non-diffractive interactions.
Diffractive interactions are characterized by the production of very few
secondaries along the direction of the incoming particles. They
correspond to the quasi-elastic exchange of a reggeon or pomeron between
virtual hadronic states the photon (mainly the vector mesons
$\rho^0$, $\omega$, and $\phi$) and the nucleon.  
Because of the spacelike nature of
 the interaction, the angular distribution is strongly forward
 peaked, and can be parametrized by Eq.~(\ref{t-dist}) with an 
energy-dependent slope 
$b_{\rm diff}= 6{\rm GeV}^{-2} + 0.5{\rm GeV}^{-2}\ln(s/s_0)$ 
\cite{Donnachie78a}.
 At high energies,
 the cross section of diffractive interactions is approximately 
a constant fraction
 of the total cross section. The relative contribution of the different
 vector mesons is predicted by theory \cite{Feynman72} to
 $\rho^0:\omega = 9:1$. The diffractive production of $\phi$ or heavier
mesons is neglected in SOPHIA.

 Our treatment of non-diffractive multiparticle production is based on the
 Dual Parton Model \cite{Capella94a}. This model can be considered as a
 phenomenological realization of the expansion of QCD for large numbers
of colors and flavours
 \cite{Hooft74,Veneziano74} in connection with general ideas of Duality and
 Gribov's Regge theory \cite{Gribov68b-e,Gribov69a-e}.  It provides a well
 developed basic scheme for the simulation of high energy hadronic
 interactions. The model can be visualized as follows: (i) the incoming
 nucleon and photon are split into colored quark and diquark constituents,
 (ii) in the course of the interaction these constituents exchange color
 quantum numbers, and (iii) confinement and the color field forces result in
 color strings which  fragment to hadrons.
 
\begin{figure}[!htb]
\begin{center}
\psfig{file=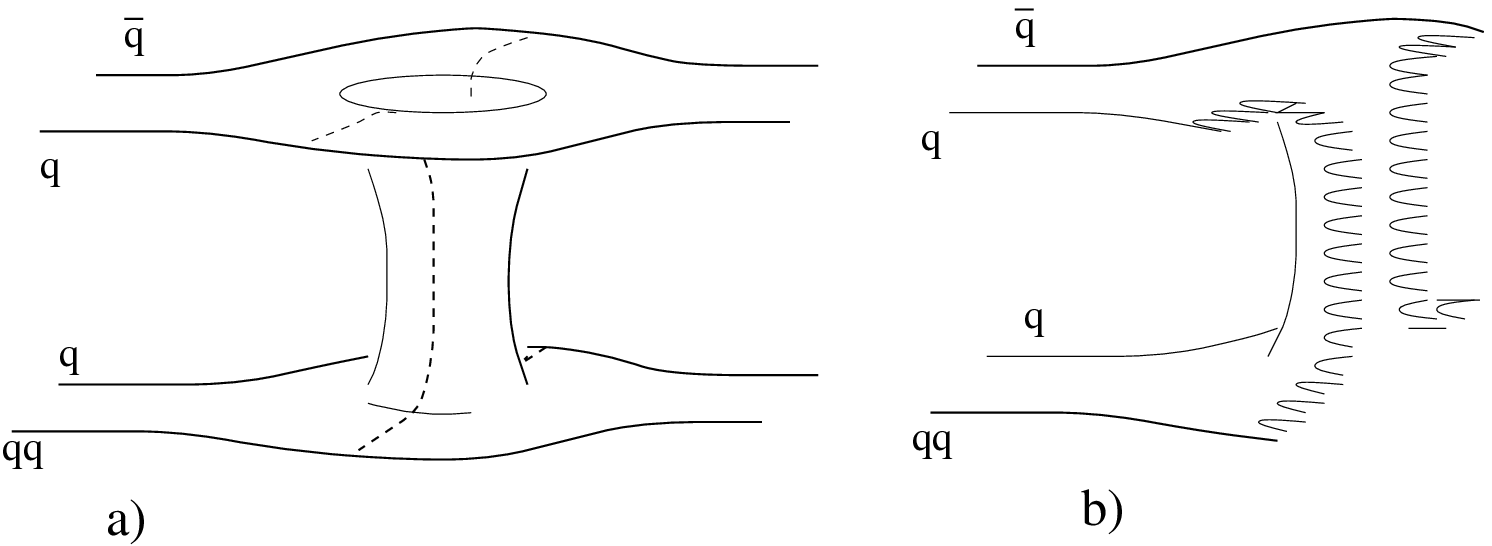,width=120mm}
\end{center}
\caption[]{\label{pcolflo}
Color flow picture of (a) a pomeron exchange graph and (b)
the final state topology}
\begin{center}
\psfig{file=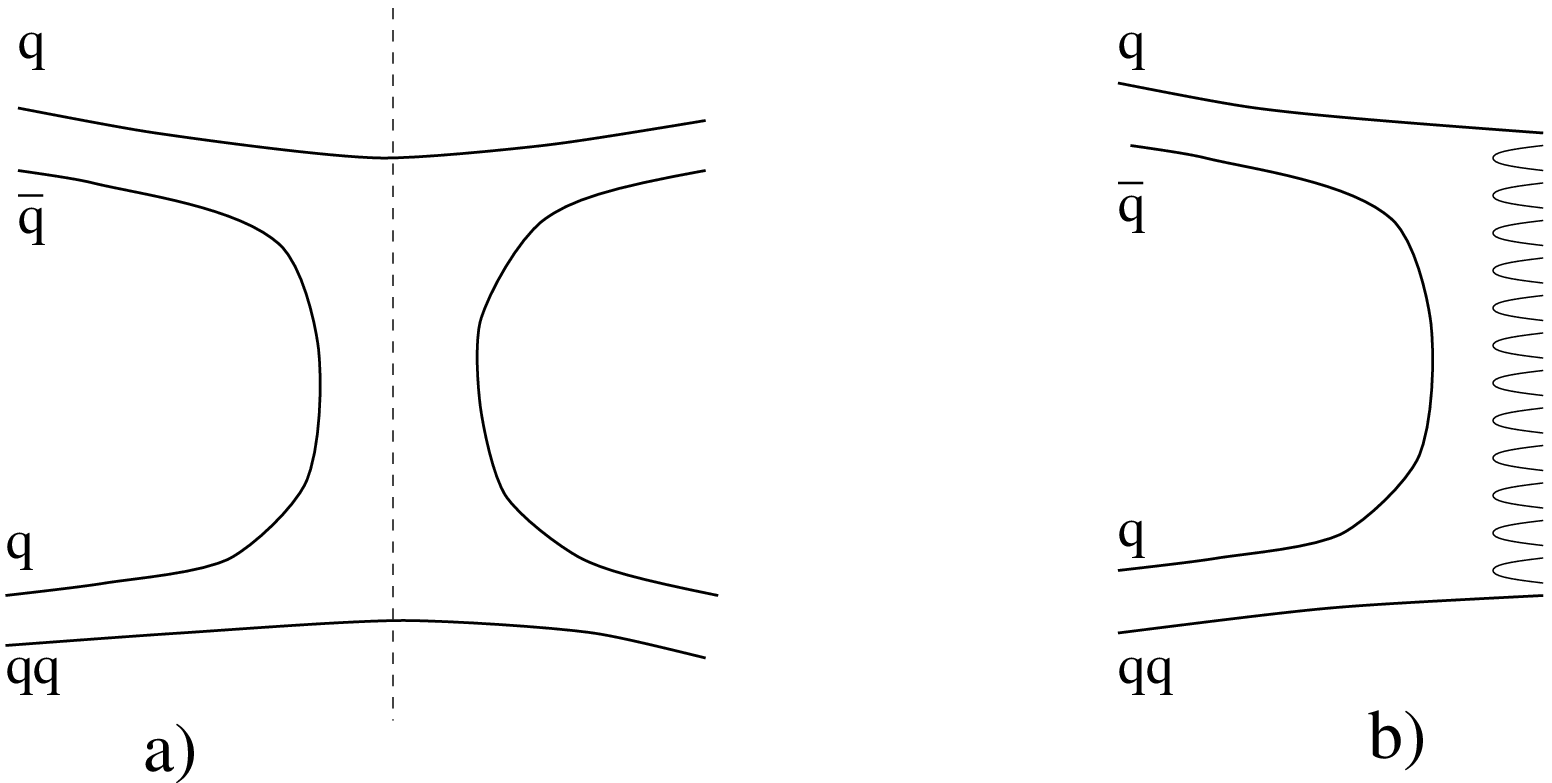,width=90mm}
\end{center}
\caption[]{\label{rcolflo}
Color flow picture of (a) a reggeon exchange graph and (b)
the final state topology}
\end{figure}

 To relate the contributions of reggeon and pomeron exchange to parton
 configurations, we use the correspondence of their respective amplitudes to
 certain color flow topologies \cite{Veneziano76} which are shown in
Figs.~\ref{pcolflo} and \ref{rcolflo}.  The pomeron exchange topology
involves the formation of two color neutral strings, while in case of a
 reggeon topology only one string is stretched from the diquark to the quark
 of the photon. The quark and diquark flavors at the string ends are
determined by the the spin and valence flavor statistics for the
nucleon.
For photons the charge difference
between $u$ and $d$ quarks increases the probability that the photon 
couples
 to a $u$-$\bar u$ pair instead of a $d$-$\bar d$ pair. In the model we use
 the theoretically predicted ratio of 4:1 between these two combinations.

 The longitudinal momentum fractions $x$, $1-x$ of the partons connected
 to the string ends are given by Regge asymptotics 
 \cite{Capella80a,Capella80b,Kaidalov82a,Kaidalov82b}.
 One gets for the valence quark ($x$) and diquark ($1-x$) distribution
 inside the nucleon
\begin{equation}
\rho(x) \sim \frac{1}{\sqrt{x}} (1-x)^{1.5}
\label{P-quark1}
\end{equation}
 and for the quark antiquark distribution inside the photon
\begin{equation}
\rho(x) \sim \frac{1}{\sqrt{x (1-x)}} .
\label{P-quark2}
\end{equation}
 The relatively small transverse momentum of the partons at the string
 ends are neglected.

In the string fragmentation process, the kinetic energy of the initial partons
is reduced by creating new quark-antiquark pairs which are
color field-connected to the parent partons. This process continues
until the available kinetic energy drops below the particle production
threshold causing the newly produced quarks to combine 
with the valence quarks to form hadrons 
(see, for example, \cite{Sjostrand88a}).

\section{Implementation}

In this section, all numerical expressions for cross sections are measured in
units of $\mubarn$, unless noted otherwise.

\subsection{Method of cross section parametrization}

The basic models of photohadronic interaction processes described in
Sect.~\ref{sect:physics} are used to obtain robust theoretical predictions
used for the parametrization of cross sections and final state distribution
functions. For theoretically unpredictable parameters we use, if possible,
the estimates given in the {\it Review of Particle Properties}
(\RPP)~\cite{Caso98a}. Remaining parameters are determined in fits to
available exclusive and inclusive data on $\gamma p$ and $\gamma n$
interactions, as compiled in standard reference series (\cite{Landolt87}, and
references therein). Since the parameters published in the \RPP\ generally
allow some variations within a given error range, these parameters are then
optimized in comparison to data in an iterative process, until a reasonable
agreement with the data for a large set of interaction channels is obtained.

This method has been previously described by Rachen~\cite{Rach96}, but was
considerably improved in the development of SOPHIA. It provides a minimum
bias description of photoproduction, which reproduces a large set of
available data while reducing a possible bias due to data selection, since
data are used only to fix model parameters. Considering the intended
applications of SOPHIA for (i) astrophysical applications and (ii) the
determination of background spectra in high energy experiments, we put
particular emphasis on a good representation of inclusive cross sections and
average final state properties in a wide range of interaction energies, while
a good representation of complex exclusive channels is generally not
expected.

\subsection{Resonance production}

Using Eq.~(\ref{breitwigner}), the contribution to the cross section from a
 resonance with mass $M$ and width $\Gamma$ can be written as a function of
 the NRF photon energy $\eps'$ as
\begin{equation}
\sigma(\eps') = \frac{s}{\epsilon^{\prime 2}} 
\frac{\sigma_0\Gamma^2s}{(s-M^2)^2+\Gamma^2s}\ .
\label{BW_impl}
\end{equation}
 The reduced cross section $\sigma_0$ is entirely determined by the resonance
 angular momentum and the electromagnetic excitation strength $b_\gamma$.  We
 selected all baryon resonances listed in the \RPP\ with certain
existence (overall status: {\tt ****})
 and a well determined minimal photo--excitation strength of
 $b_\gamma > 10^{-4}$ for either the $p\gamma$ or the $n\gamma$
 excitation. The resonances fulfilling these criteria and their parameters,
 as implemented in SOPHIA after iterative optimization, are given in
 Table~\ref{tab:res}. The phase-space reduction close to the $N\pi$ threshold
 is heuristically taken into account by multiplying Eq.~(\ref{BW_impl}) with
 the linear quenching function $\Qf(\eps';0.152,0.17)$ for the
 $\Delta(1232)$-resonance, and with $\Qf(\eps';0.152,0.38)$ for all other
 resonances. The function $\Qf(\eps';\eps'_{\rm th},w)$ is defined in
 Appendix~\ref{app:functions}. The quenching width $w$ has been determined
 from comparison with the data of the total $p\gamma$ cross section, and of
 the exclusive channels $p\pi^0$, $n\pi^+$ and $\Delta^{++}\pi^-$ where most
 of the resonances contribute.
\begin{table}
\centering
\caption[]{\label{tab:res} Baryon resonances and their physical parameters
implemented in SOPHIA (see text). Superscripts $^+$ and $^0$ in the
parameters refer to $p\gamma$ and $n\gamma$ excitations, respectively. The
maximum cross section, $\sigma_{\rm max} = 4 m_N^2 M^2 \sigma_0/(M^2 -
m_N^2)^2$, is also given for reference.}
\begin{center}
\renewcommand{\arraystretch}{1.2}
\begin{tabular}{|cll|rrr|rrr|}\hline
resonance & \cc{$M$} & \ccl{$\Gamma$} & \cc{$10^3 b_\gamma^+$} & 
\cc{$\sigma_0^+$} & \ccl{$\sigma_{\rm max}^+$} & \cc{$10^3 b_\gamma^0$} & 
\cc{$\sigma_0^0$} & \ccl{$\sigma_{\rm max}^0$} \\\hline
$\Delta(1232)$ 	& 1.231 &  0.11  & 5.6 & 31.125 & 411.988 & 6.1 & 33.809 & 
452.226\\
$N(1440)$  	& 1.440 &  0.35  & 0.5 &  1.389 &   7.124 & 0.3 &  0.831 &
   4.292\\
$N(1520)$  	& 1.515 &  0.11  & 4.6 & 25.567 & 103.240 & 4.0 & 22.170 &
  90.082\\
$N(1535)$  	& 1.525 &  0.10  & 2.5 &  6.948 &  27.244 & 2.5 &  6.928 &
  27.334\\
$N(1650)$  	& 1.675 &  0.16  & 1.0 &  2.779 &   7.408 & 0.0 &  0.000 &
   0.000\\
$N(1675)$  	& 1.675 &  0.15  & 0.0 &  0.000 &   0.000 & 0.2 &  1.663 &
   4.457\\
$N(1680)$  	& 1.680 &  0.125 & 2.1 & 17.508 &  46.143 & 0.0 &  0.000 &
   0.000\\
$\Delta(1700)$  & 1.690 &  0.29  & 2.0 & 11.116 &  28.644 & 2.0 & 11.085 &
  28.714\\
$\Delta(1905)$  & 1.895 &  0.35  & 0.2 &  1.667 &   2.869 & 0.2 &  1.663 &
   2.875\\
$\Delta(1950)$  & 1.950 &  0.30  & 1.0 & 11.116 &  17.433 & 1.0 & 11.085 &
  17.462\\
\hline
\end{tabular}
\end{center}
\end{table}
 The major hadronic decay channels of these baryon resonances are $N\pi$,
 $\Delta\pi$, and $N\rho$; for the $N(1535)$, there is also a strong decay
 into $N\eta$, and the $N(1650)$ contributes to the $\Lambda K$ channel. The
 hadronic decay branching ratios $b_c$ are all well determined for these
 resonances and given in the \RPP. However, a difficulty arises from the
 fact that branching ratios can be expected to be energy dependent because
 of the different masses of the decay products in different branches.
 In SOPHIA, we consider all secondary particles, including hadronic resonances,
 as particles of a fixed mass. This implies that, for example, the decay 
channel $\Delta\pi$ is energetically forbidden for
 $\sqrt{s} < m_\Delta+m_\pi \approx 1.37$ GeV. To accommodate this problem,
 we have developed a scheme of energy dependent branching ratios, which
 change at the thresholds for additional decay channels and are constant
 in between. The requirements are that (i) the branching ratio $b_c=0$ for
 $\eps'<\eps'_{{\rm th},c}$, and (ii) the average of the branching ratio over
 energy, weighted with the Breit-Wigner function, correspond to the average
 branching ratio given in the \RPP\  for this channel. For all resonances,
 we considered not more than three decay channels leading to a unique
 solution to this scheme. No fits to data are required.  In practice,
 however, the experimental error on many branching ratios allows for
some freedom, which we have used to generate a scheme that optimizes 
the agreement with the data on different exclusive channels.

 The hadronic branching ratios are given in Tab.~\ref{tab:resbranch}
 in Appendix~\ref{app:branches}.  To obtain the contribution to a
 channel with given particle charges, e.g. $\Delta^{++}\pi^-$, the
 hadronic branching ratio $b_{\Delta\pi}$ has to be multiplied with
 the iso-branching ratios as given in Tab.~\ref{tab:iso}. We note
 that with the parameters $b_\gamma$, $b_c$ and $b_{\rm iso}$, the
 resonant contribution to all exclusive decay channels is completely
 determined.

The angular decay distributions for the resonances follow from
Eq.~(\ref{decay-distributions}). In SOPHIA, the kinematics of
the decay channels into
$N\pi$ is implemented in full detail (see Tab.~\ref{tab:resang}).
For other decay channels, we assume isotropic decay according to the
phase space. Furthermore, there might be
some mixing of the different scattering angular distributions since 
the sampled resonance mass, in general, does not coincide with its nominal 
mass.   This effect is
neglected in our work. Instead, 
we use the angular distributions applying to resonance decay at 
its nominal mass $M$.
\begin{table}[!htb]
\caption[]{\label{tab:resang}
Angular distribution probability functions for $N\pi$ decay of resonances
considered in SOPHIA. The resonances N(1535), N(1650) and N(1440) decay
isotropically. 
}  
\begin{center}
\renewcommand{\arraystretch}{1.2}
\begin{tabular}{|c|l|}
\hline
resonance & $ \cP(\cos\chi^*)$  \\\hline
$\Delta(1232)$ & $0.636263 - 0.408790\cos^2\chi^*$\\
$N^0(1520)$    & $0.673669 - 0.521007\cos^2\chi^*$\\
$N^+(1520)$    & $0.739763 - 0.719288\cos^2\chi^*$\\
$N^0(1675)$    & $0.254005 + 1.427918\cos^2\chi^* - 1.149888\cos^4\chi^*$\\ 
$N^+(1680)$    & $0.189855 + 2.582610\cos^2\chi^* - 2.753625\cos^4\chi^*$\\ 
$\Delta(1700)$ & $0.450238 + 0.149285\cos^2\chi^*$\\
$\Delta(1905)$ & $0.230034 + 1.859396\cos^2\chi^* - 1.749161\cos^4\chi^*$\\
$\Delta(1950)$ & $0.397430 - 1.498240\cos^2\chi^* + 5.880814\cos^4\chi^* 
			- 4.019252\cos^6\chi^*$\\\hline
\end{tabular}
\end{center}
\end{table}

The two decay products of a resonance may also decay subsequently.
This decay is simulated to occur isotropically according to the 
available phase space.

\subsection{Direct pion production}

 The cross section for direct meson production, unlike those of resonances, is
 not completely determined by well known parameters. The low and high energy
 constraints suggest the phenomenological parametrization
\begin{equation}
\label{sigmadir}
\sigma_{\rm dir}(\eps') = \sigma_{\rm max} 
	\Pl(\eps';\eps'_{\rm th}, \eps'_{\rm max}, 2)\;,
\end{equation}
 where the function $\Pl(\eps';\eps'_{\rm th}, \eps'_{\rm max}, \alpha)$
 approaches zero for $\eps'=\eps'_{\rm th}$, goes through a maximum at $\eps'
 = \eps'_{\rm max}$ and follows an asymptotic behaviour $\propto
 (\eps')^{-\alpha}$. The definition of this function is given in Appendix
 \ref{app:functions}.  

In SOPHIA, we consider explicitly direct
channels with charged pion exchange which are dominating at low
energies. The selection is further constrained
by the fact that sufficient data are only available for the channels $p\gamma
\to n \pi^+$, $n\gamma\to p \pi^-$, and $p\gamma \to \Delta^{++}\pi^-$.
We note that
proton and neutron induced direct reactions are strictly
 isospin-symmetric. Both proton and neutron data sets (when available) 
can be used in the fitting procedure. The high energy data
  fits on the $\Delta\pi$ and
 $N\pi$ channel, i.e. $\sigma_{\pi} \approx 18(\eps')^{-2}$ \cite{boya68} and
 $\sigma_\Delta\approx 26.4 (\eps')^{-2}$ \cite{boya69} 
for $\eps'>1$, were primarily
 used to fix $\sigma_{\rm max}$, while a best fit of $\eps'_{\rm{max}}$ was
 obtained by comparing with the residuals of the low energy data after
 subtracting the resonance contribution. The adopted cross sections are
\begin{eqnarray}
\label{sigma:dir:Npi}
\sigma_{N\pi}(\eps') &=& 92.7 \Pl(\eps';0.152,0.25,2)+40\exp\left(
-\frac{(\eps'-0.29)^2}{0.002}\right)\\
& & -15\exp\left(-\frac{(\eps'-0.37)^2}{0.002}\right)\ ,
\nonumber\\
\label{sigma:dir:Dpi}
\sigma_{\Delta\pi}(\eps') &=& 37.7 \Pl(\eps';0.4,0.6,2)\ .
\end{eqnarray}
 The two Gaussian-shaped functions included in the direct $N\pi$ cross
 section have been added to improve the representation of the total cross
 section in the energy region $0.152$~GeV~$<\eps'<0.4$~GeV, 
where otherwise only
 the well constrained $\Delta(1232)$ resonance contributes significantly.  For
 $p\gamma$- ($n\gamma$-) interactions $\sigma_{\Delta\pi}$ contributes to the
 $\Delta^{++}\pi^-$ ($\Delta^{+}\pi^-$) and $\Delta^0\pi^+$ ($\Delta^-\pi^+$)
 final states with a ratio 3:1 according to isospin combination rules (see
Tab.~\ref{tab:iso}).

 By comparison with the total cross section data we find that the
 resonant and direct interaction channels account for all of the total
 interaction cross section below the $3\pi$ threshold at $\eps'\approx
 0.51$ GeV. Above this threshold, and below the threshold for diffractive
 interactions at $\eps'\approx 1$ GeV, where high energy processes set in,
 we find a residual cross section which can be parametrized as  
\begin{equation}
\sigma_{\rm lf} = 80.3 (60.2) \Qf(x;0.51;0.1) (\eps')^{-0.34} \ ,
\end{equation}
 where the number 60.2 given in brackets belongs to $n\gamma$--interactions
while the number 80.3 refers to $p\gamma$--collisions. The normalization cross
section and the quenching width has been determined by a $\chi^2$
minimization method to the total cross section data for $p\gamma$ ($n\gamma$)
interactions after subtraction of the respective resonant and direct
contributions. By analogy, the power law index for this contribution is taken
from the high energy parametrization for reggeon exchange (note that
$\eps'\propto s-m_N^2$). Physically, this cross section represents the joint
contribution of all $t$-channel scattering processes at low energies not
considered so far. This is in principle similar to interactions at high
energies. Consequently, we use an adapted string fragmentation model to
simulate this contribution, and refer to it as {\it low energy fragmentation}
hereafter.

 \subsection{High energy multipion production}

 In SOPHIA, we assume that the cross sections for diffractive and
 non-diffractive high energy interactions are proportional to each
 other at all energies. This assumption fixes the threshold for
 high energy interactions to the threshold of the $N\rho$ final state,
 which is nominally at $\eps'\approx1.1$ GeV. Because of the large width
 of the $\rho$ there should be some contribution also at lower energies.
 From comparison with exclusive data of the $N\rho$ final state,
 and the residuals of the total cross section data, with the sum of the
 contributions of all low energy channels, we find a common threshold
 for high energy interactions of $\eps'_{\rm th,high} = 0.85$ GeV. 

We restrict the diffractive channel to the non-resonant production of
$N\rho$ and $N\omega$ final states, for which we assume the theoretically
 predicted relation $\sigma_\rho = 9\sigma_\omega$.
The ratio between diffractive and
non-diffractive interactions is derived from the comparison with exclusive
$N\rho$ data and total cross section data at high energy,
\begin{equation}
\sigma_{\rm diff} = 0.15\ \sigma_{\rm frag} \;.
\end{equation}
 For the parametrization of $\sigma_{\rm frag}$, we use the power law
 representations of the reggeon and pomeron exchange cross section at high
 energies, and multiply them by an exponential quenching function
$1-\exp([\eps'_{\rm th,high}-\eps']/a)$. 
The relative contributions of the reggeon and pomeron cross
 sections, and the quenching parameter $a$ have been determined by a
 iterative $\chi^2$ minimization method with respect to the
 total $p\gamma$ ($n\gamma$) cross section data after subtraction of all low
 energy contributions. We find
\begin{equation}
\sigma_{\rm{frag}}(\eps') = \left[1-\exp\left(-\frac{\eps'-0.85}{0.69}\right)
	\right][28.8 (26.0)(\eps')^{-0.34} + 58.3 (\eps')^{0.095}]\;,
\end{equation}
where we have used the high-energy behaviour given by Eq.~(\ref{DL-model}).

The string fragmentation is done by the Lund Monte Carlo {\sc Jetset} 7.4
\cite{Sjostrand88a,Sjostrand94a}.  This program is well suited for string
fragmentation at high energies.  Since for our purposes also strings with
rather small invariant masses have to be hadronized, several parameters of
this fragmentation code had to be tuned to obtain a reasonable description
also at low energies. Furthermore, in order to avoid double counting, all
final states identical to the processes already considered by resonance
production and direct interactions are rejected (note that this is not the
case for low-energy fragmentation).

\subsection{Initial state kinematics and photon radiation fields}

The probability for interaction of a proton of energy $E_N$ with a photon of
energy $\epsilon$ from a radiation field with the photon density
$n(\epsilon)$ reads
\begin{equation}
\cP(\epsilon) = \frac{1}{R(E_N)}\,\frac{n(\epsilon)}{8 E_N^2 \beta \epsilon^2} 
\int_{s_{\rm th}}^{s_{\rm max}} ds
(s-m_N^2) \sigma_{N\gamma}(s)\ ,
\end{equation}
where $R(E_N)$ is the interaction rate as given in Eq.~(\ref{intrate}), where
also $s_{\rm th}$ and $s_{\rm max}$ are defined.

For a fixed nucleon energy, the CMF energy is sampled from the distribution
\begin{equation}
\cP(s) = \Phi^{-1} (s-m_N^2) \sigma_{N\gamma}(s)
\end{equation}
with $\Phi = \int_{s_{\rm th}}^{s_{\rm max}} ds
(s-m_N^2) \sigma_{N\gamma}(s)$.
The interaction angle follows from
\begin{equation}
\cos{\theta} = \frac{1}{\beta}
\left(\frac{m_N^2-s}{2E_N\epsilon}+1\right) .
\end{equation}

Currently black body, power law, and broken power law radiation spectra are
implemented in SOPHIA.  The photon density $n(\epsilon)$ for a blackbody
radiation field of temperature $T$ is given in natural units by
\begin{equation}
n(\epsilon) = \frac1{\pi^2}
\frac{\epsilon^2}{\exp(\frac{\epsilon}{k T})-1},
\end{equation}
where $k$ is the Boltzmann constant.  For a power law photon spectrum the
photon density is given by $n(\epsilon) = \epsilon^{-\alpha}$. The broken
power law photon spectrum is given by
\begin{eqnarray}
n(\epsilon) =& \epsilon^{-\alpha_1} & \mbox{for}\;\;
 \epsilon<\epsilon_b\\
n(\epsilon) =& \epsilon_b^{\alpha_2 - \alpha_1} \epsilon^{-\alpha_2} 
 & \mbox{for} \hspace*{.2cm} 
\epsilon>\epsilon_b
\end{eqnarray}
where $\epsilon_b$ is the break energy, and $\alpha_1$, $\alpha_2$ are the
photon indices below and above the break energy, respectively.  Note that no
absolute normalization of $n(\eps)$ is necessary since it cancels in the
definition of $\cP(\eps)$.

\section{Structure of the program}

\subsection{Source code}

 The SOPHIA source code consists of several files which contain a number
of routines.
\begin{description}
\item[sophia.f] main program containing the routines which organize
 the various tasks to be performed. Furthermore, the input is
 handled here and some kinematic transformations needed as input
 to several routines are performed.
\item[initial.f] initialization routine for parameter settings.
\item[sampling.f] collection of routines/functions needed for sampling
 the CMF energy squared $s$ and the photon energy $\epsilon$ in the
 observer frame. 
\item[eventgen.f] event generator for photomeson production in $p\gamma$
and $n\gamma$ collisions.
\item[output.f] contains output routines.
\end{description}

\subsection{The event generator}

The simulation of the final state is performed by the photopion
production event generator {\tt EVENTGEN}.  Together with the
initialization routine {\tt INITIAL}, {\tt EVENTGEN} can be 
used separately for Monte Carlo event simulation. The user has to give
the nucleon code number {\tt L0}, energy {\tt E0} (in GeV),
the photon of energy $\epsilon$ (in GeV), and the angle $\theta$ in
degrees.

{\tt EVENTGEN} 
is structured as follows. First the momenta of the incident particles
are  Lorentz-transformed into the CMF of the interactions.
 The cross section (in $\mu$barn) is calculated in the function 
 {\tt CROSSECTION}. The cross sections for the various channels considered
 in this code determine the distribution for the probability of
 a certain process. The sampling of a process (resonance decay, direct channel,
 diffractive scattering, fragmentation) at a given NRF energy
 of the photon is carried out in the routine {\tt DEC\_INTER3}.

 For the resonance decay we sample the resonance at this energy using the
 Breit-Wigner formula as a probability distribution for a specific resonance
 (performed in {\tt DEC\_RES2}). Its branching
ratios define the decay mode (in {\tt DEC\_PROC2}). The subsequent two-particle
  decay in the CM frame
is carried out in {\tt RES\_DECAY3}, and then decays of all
 unstable particles are carried out in the {\sc Sybill} routine {\tt DECSIB}.

Secondary particle production is simulated in 
{\tt GAMMA\_H} for
direct and multiparticle production. 

Finally, all final state particles are Lorentz-transformed
back to the LF. 

The output of the final states is organized in the
common block {\tt /S\_PLIST/ P(2000,5), LLIST(2000), NP, Ideb}. Here the array
 {\tt P(i,j)} contains the 4-momenta and rest mass of the final state 
particle {\tt i}
in cartesian coordinates 
({\tt P(i,1)} = $P_x$, {\tt P(i,2)} = $P_y$, {\tt P(i,3)} = $P_z$,
{\tt P(i,4)} = energy, {\tt P(i,5)} = rest mass).
{\tt LLIST()} gives the code numbers of all final state particles and
{\tt NP} is the number of stable final state particles.

\subsection{Input/Output routines}

Using the standard main program, the user provides the following
input parameters.
\begin{itemize}
\item {\tt E0} = energy of incident proton (in GeV), or
\item {\tt Emin, Emax} = low/high energy cutoff of an energy grid of 
incident protons (in GeV)
\item {\tt L0} = code number of the incident nucleon 
({\tt L0} = 13: proton, {\tt L0} = 14: neutron)
\item ambient photon field:   
\begin{itemize}
\item blackbody spectrum: {\tt TBB} = temperature (in K) 
\item straight/broken power law spectrum:  {\tt ALPHA1}, 
{\tt ALPHA2} = power law
 indices, {\tt EPSMIN} = low energy cut off  (in eV), 
{\tt EPSMAX} = high energy cut off (in eV), {\tt EPSB} = break energy (in eV)
\end{itemize}
\item {\tt NTRIAL} 
= number of inelastic interactions
\item {\tt NBINS} 
= number of bins for output particle spectra
 ($\le 200$ bins)
\item {\tt DELX} = stepsize of output particle spectra
\end{itemize}
For the calculation of the incident particle momenta we assume that the
 relativistic nucleon
is moving along the positive $z$-axis.

The output is organized as follows.
All the energy distributions $(1/N_{\rm eve}) dN_{\rm part}/d\log{x}$ 
of produced
particles are given with logarithmically equal 
bin sizes in the scaling variable $x=E_{\rm part}$/{\tt E0}. 
Here $N_{\rm eve}$ denotes the number of simulated 
inelastic events and $N_{\rm part}$ is the number of secondary particles of
a certain kind. The spectra of photons, protons, neutrons,
$e$-neutrinos, $e$-antineutrinos, $\mu$-neutrinos and $\mu$-antineutrinos are 
considered separately.
They are stored in a file with name xxxxxx.particle
with xxxxxx = input name (chosen by the user):

\begin{tabular}{ccccc}
     particle = & 'gamma' & $\longrightarrow$ & $\gamma$       & spectrum\\
                & 'muneu' & $\longrightarrow$ & $\nu_\mu$      & spectrum\\
                & 'muane' & $\longrightarrow$ & $\bar\nu_\mu$  & spectrum\\
                & 'e\_neu' & $\longrightarrow$ & $\nu_e$       & spectrum\\
                & 'eaneu' & $\longrightarrow$ & $\bar\nu_e$    & spectrum\\
                & 'elect' & $\longrightarrow$ & $e^-$          & spectrum\\
                & 'posit' & $\longrightarrow$ & $e^+$          & spectrum\\
                & 'proto' & $\longrightarrow$ & $p$            & spectrum\\
                & 'neutr' & $\longrightarrow$ & $n$            & spectrum\\
\end{tabular}

The structure of a typical output file is:\\

\begin{tabular}{ll}
1.~line: & high/low energy cutoff of incident nucleon energy grid,\\
        & number of energy bins of incident nucleon spectrum 
(={\tt NINC}),\\
        & {\tt TBB} or {\tt ALPHA1}, 
{\tt ALPHA2}, {\tt EPSMIN}, {\tt EPSB}, {\tt EPSMAX}, incident particles\\
\end{tabular}

\begin{tabular}{lll}
2.~line: & 1.~number: & energy of incident nucleon\\
        & 2.~number: & first number (=$a$) of non-zero energy bin of\\
        &           & particle spectrum\\
        & 3.~number: & last number (=$b$) of non-zero energy bin of\\
        &           & particle spectrum\\
3.~line: & particle spectrum & \\
        & between $a$ \ldots $b$ & \\
\end{tabular}

\section{Comparison to data}

\subsection{Total cross section}

\begin{figure}
\vbox{\psfig{figure=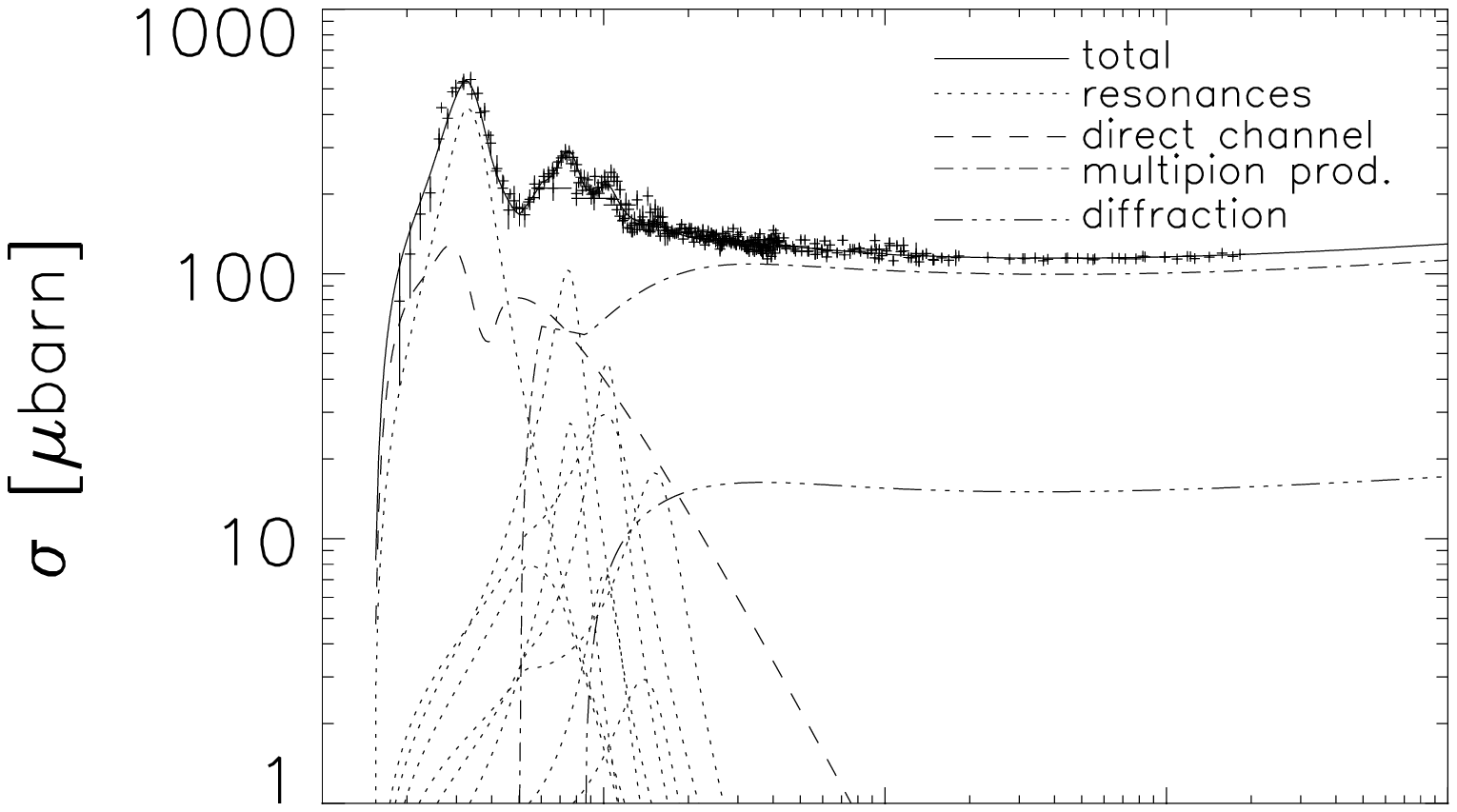,height=9cm,width=16cm}}
\vspace*{-0.45cm}
\vbox{\psfig{figure=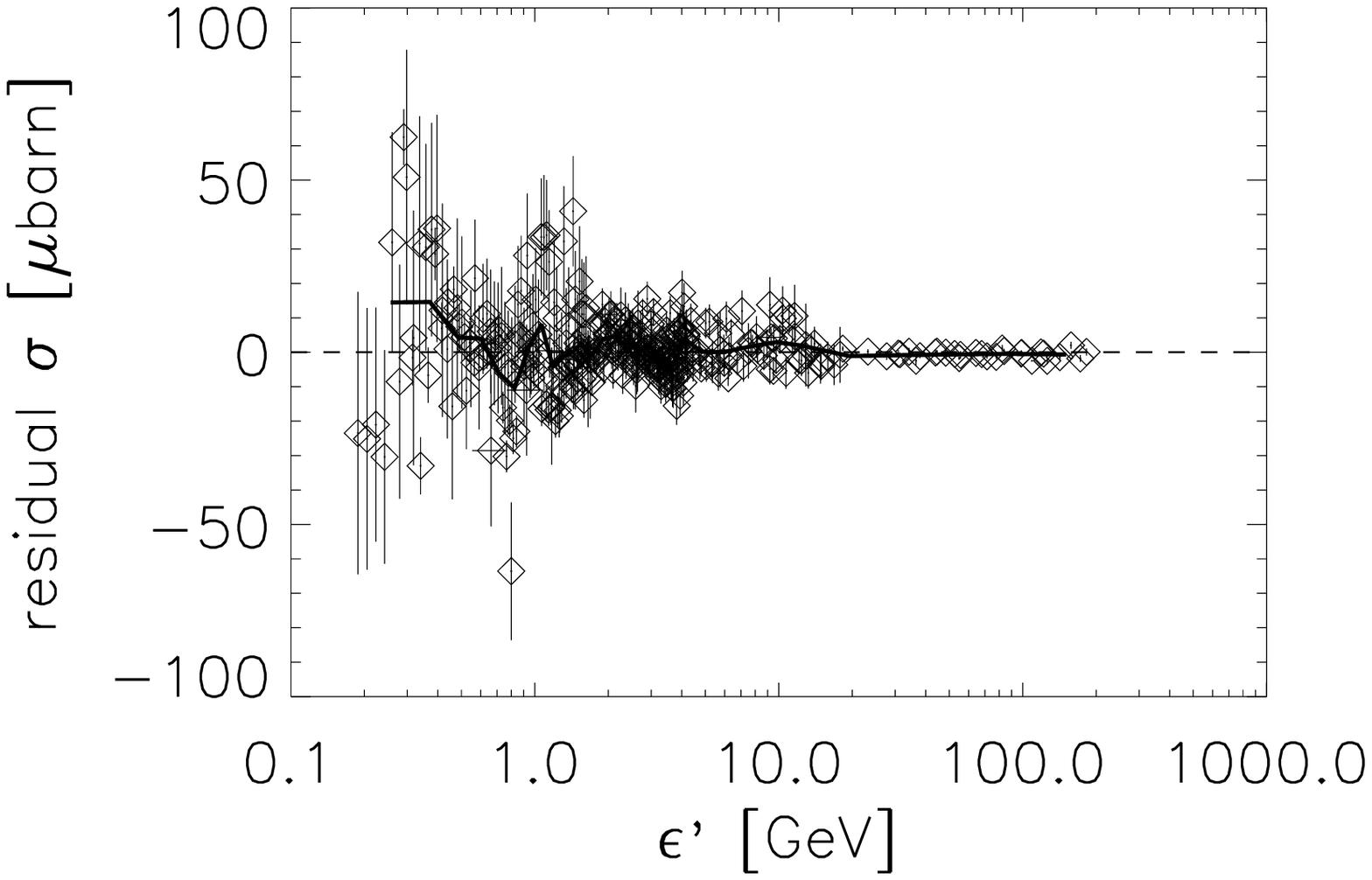,height=7cm,width=16cm}}
\caption[fig1]{
\label{sigma-tot}
The total $\gamma p$ cross section (solid line) with the contributions of
baryon resonances (dotted lines), 
direct pion production processes (dashed line)
non-diffractive multipion production (dash-dotted line), and diffractive
scattering (lower dash-triple-dotted line). Data are taken from 
\cite{arms,ball69a,meye,diet,cald,mich,alex,bing73}.  Bottom panel:
Residuals of the total cross section data to the sum of all partial cross
sections implemented in SOPHIA; the line shows the average over 10
neighboring data points.}
\end{figure}
Fig.~\ref{sigma-tot} (upper panel) shows the total cross section for $\gamma
p$-interactions with the various contributions considered in SOPHIA. For
simplicity, we show both fragmentation contributions together (low energy
fragmentation and non-diffractive multipion production).  The resonant
process $\gamma p \rightarrow \Delta^+ \rightarrow \pi^0 p$ is the only one
kinematically possible directly at the particle production threshold.  Above
the $\pi^+ n$ threshold at very low energies ($\eps'<0.25\GeV$), the direct
channel $\gamma p \rightarrow \pi^+ n$ constitutes the largest contribution.

To assess the quality of the cross section parametrization, the
 differences between the experimental data on the total $\gamma p$ cross
 section and the cross section fit are shown in Fig.~\ref{sigma-tot} 
(lower panel).  The
 total $\gamma n$ cross section is overall similar, except at energies of
 about $\sqrt{s}\approx 1.680$ GeV ($\epsilon^\prime \approx 1.035$ GeV),
where
it is considerably smaller due to the different excitation strengths of the
resonances at this energy.

\subsection{Exclusive cross sections}	

\begin{figure}[!htb]
{\psfig{figure=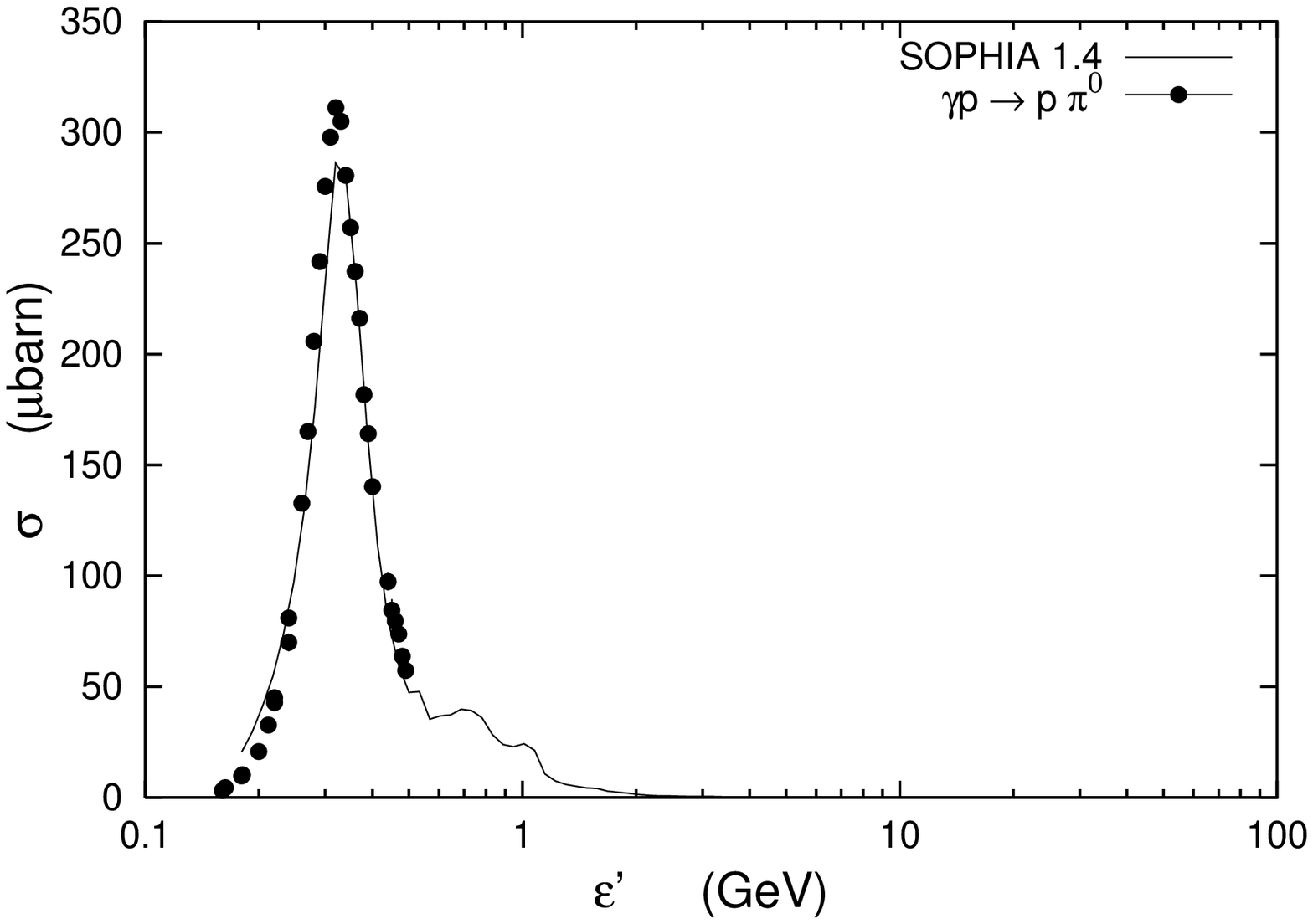,width=14cm,height=8.cm}}
{\psfig{figure=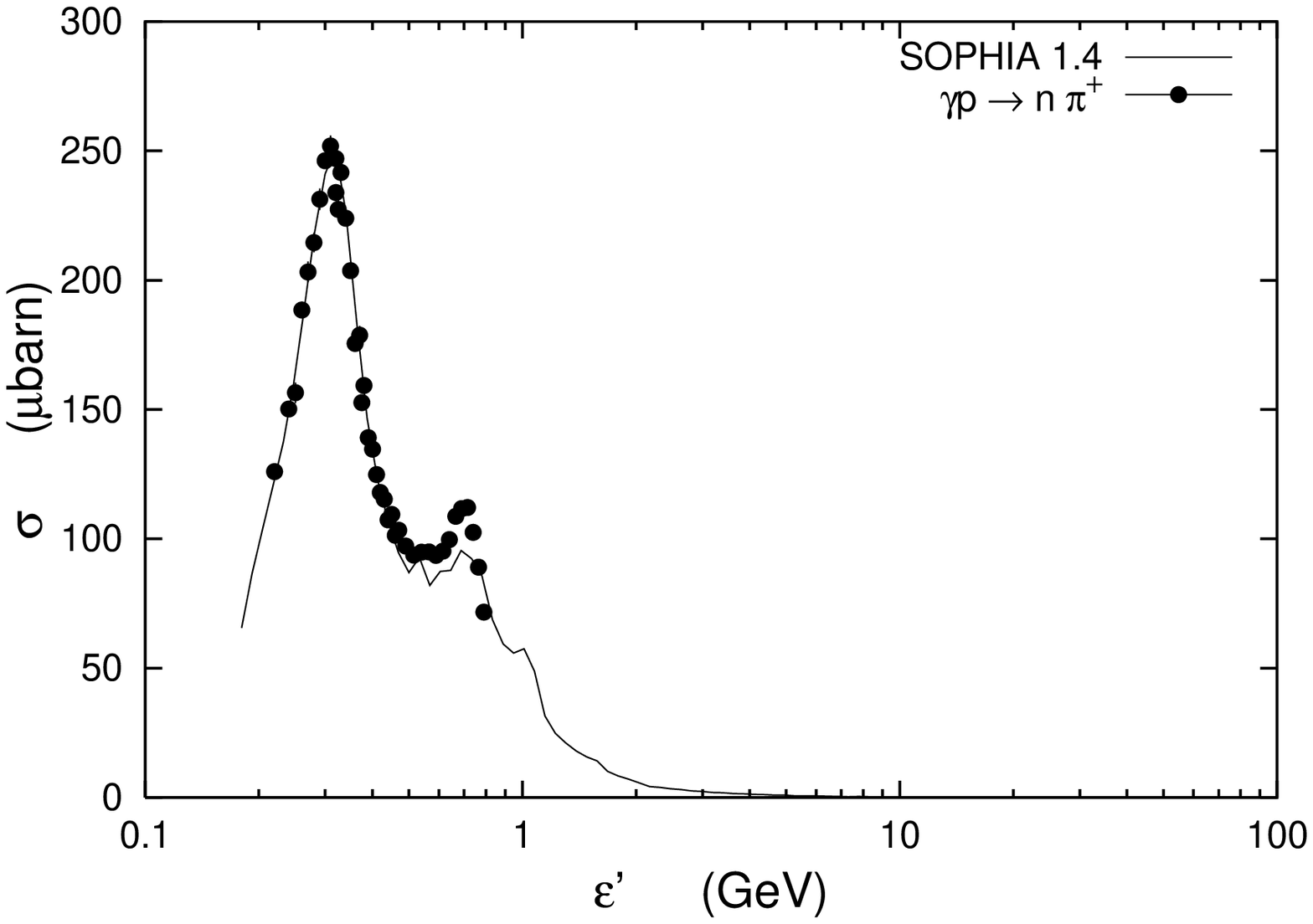,width=14cm,height=8.cm}}
\caption[fig1]{
\label{pion-spec-0}
Total cross section of 
$\gamma p\longrightarrow \pi^0 p$ and
$\gamma p\longrightarrow  \pi^+ n$. Data are from 
\cite{vasi60,govo67,fisc72a,fisc72b,genz74,fuji77}.}
\end{figure}

\begin{figure}[!htb]
{\psfig{figure=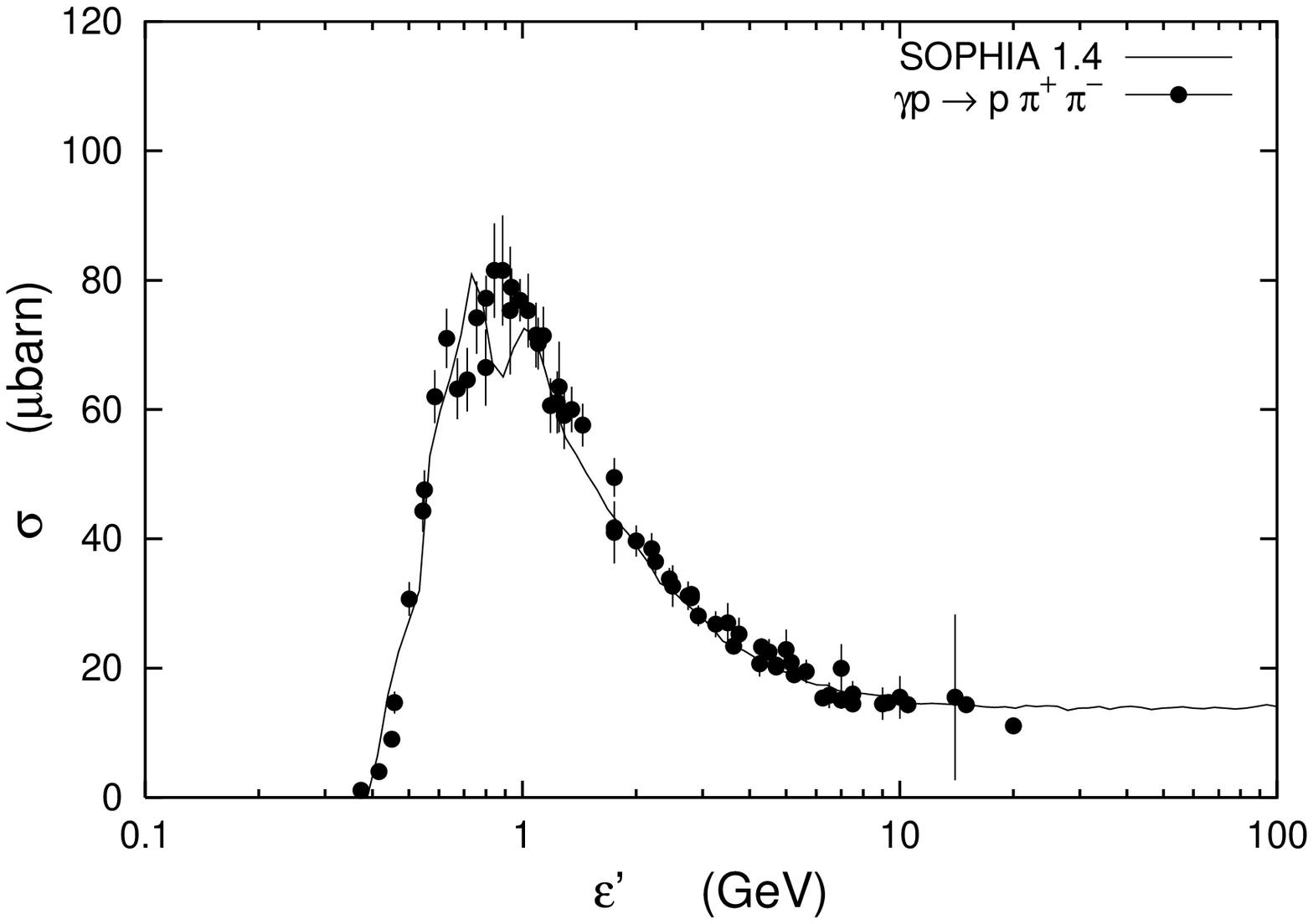,width=14cm,height=6.5cm}}
{\psfig{figure=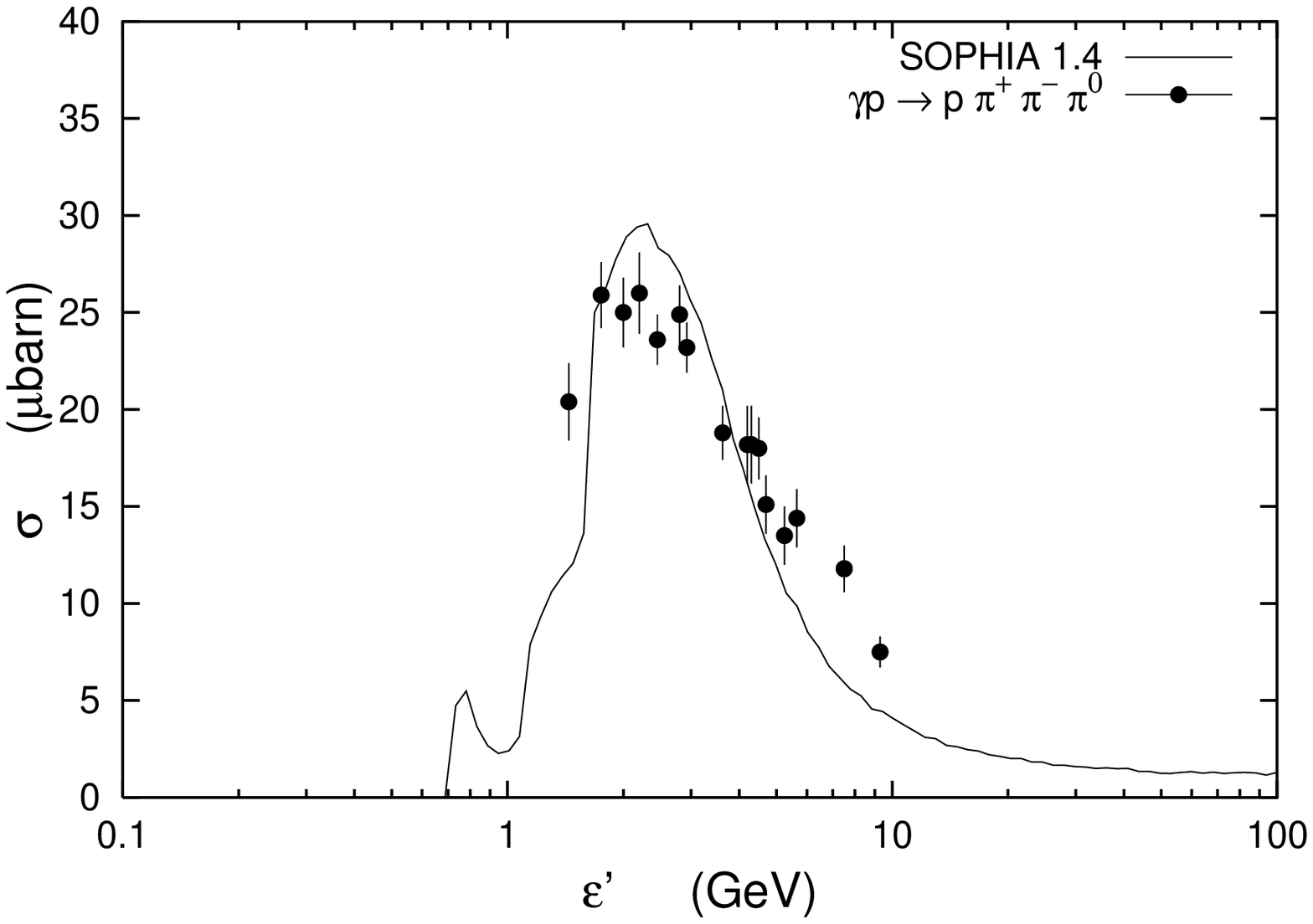,width=14cm,height=6.5cm}}
{\psfig{figure=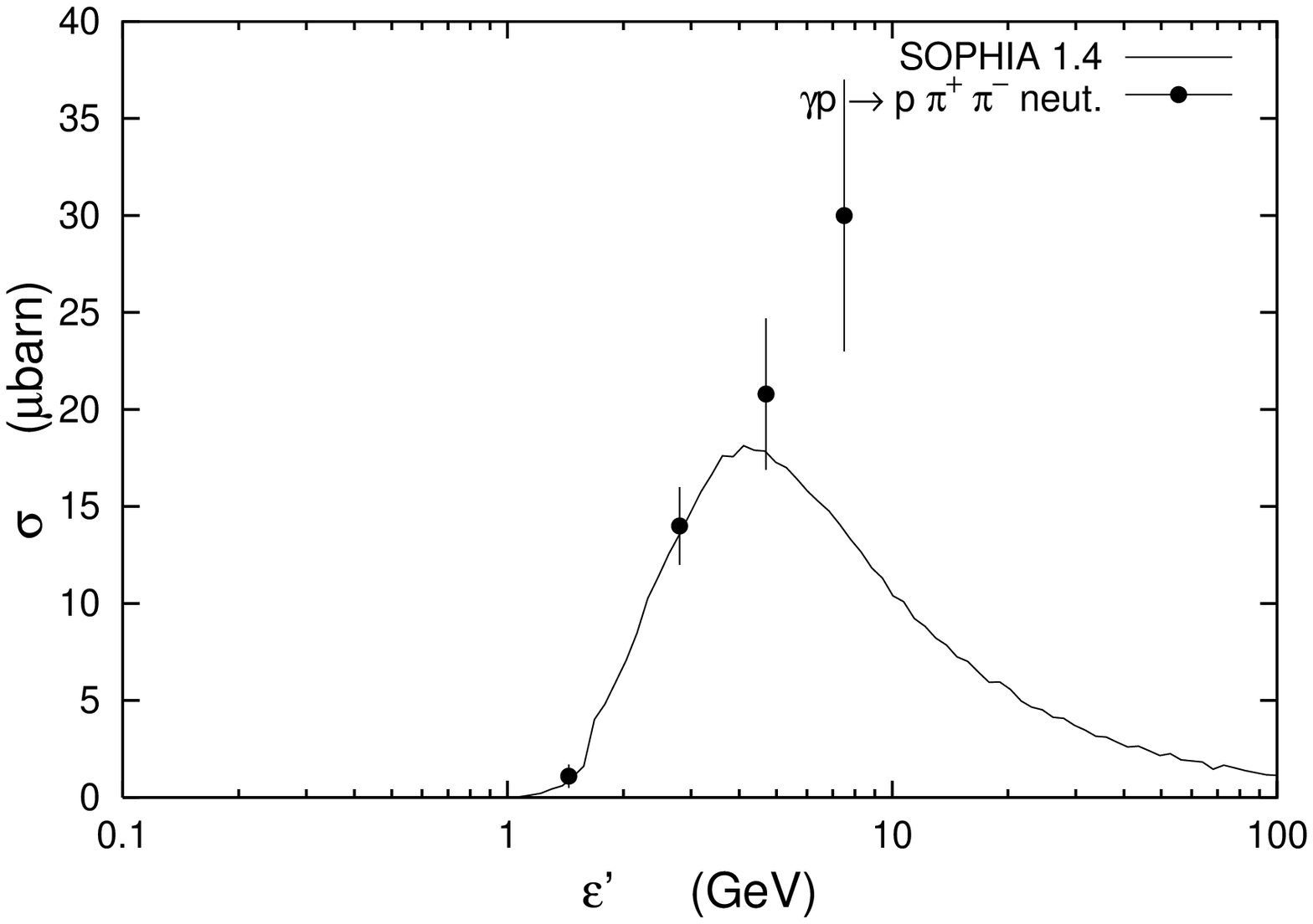,width=14cm,height=6.5cm}}
\caption[fig1]{
\label{pion-spec-1}
Total cross section of 
$\gamma p\longrightarrow \pi^+\pi^- p$, 
$\gamma p\longrightarrow \pi^+\pi^-\pi^0 p$ and 
$\gamma p\longrightarrow \pi^+\pi^- p+ ${\it neutrals}. Data are from 
\cite{ball68,stru76,ball72,ball74,carb76,haus,eise72,davi68,bing73,park72,%
abe84,alex65}.}
\end{figure}

\begin{figure}[!htb]
{\psfig{figure=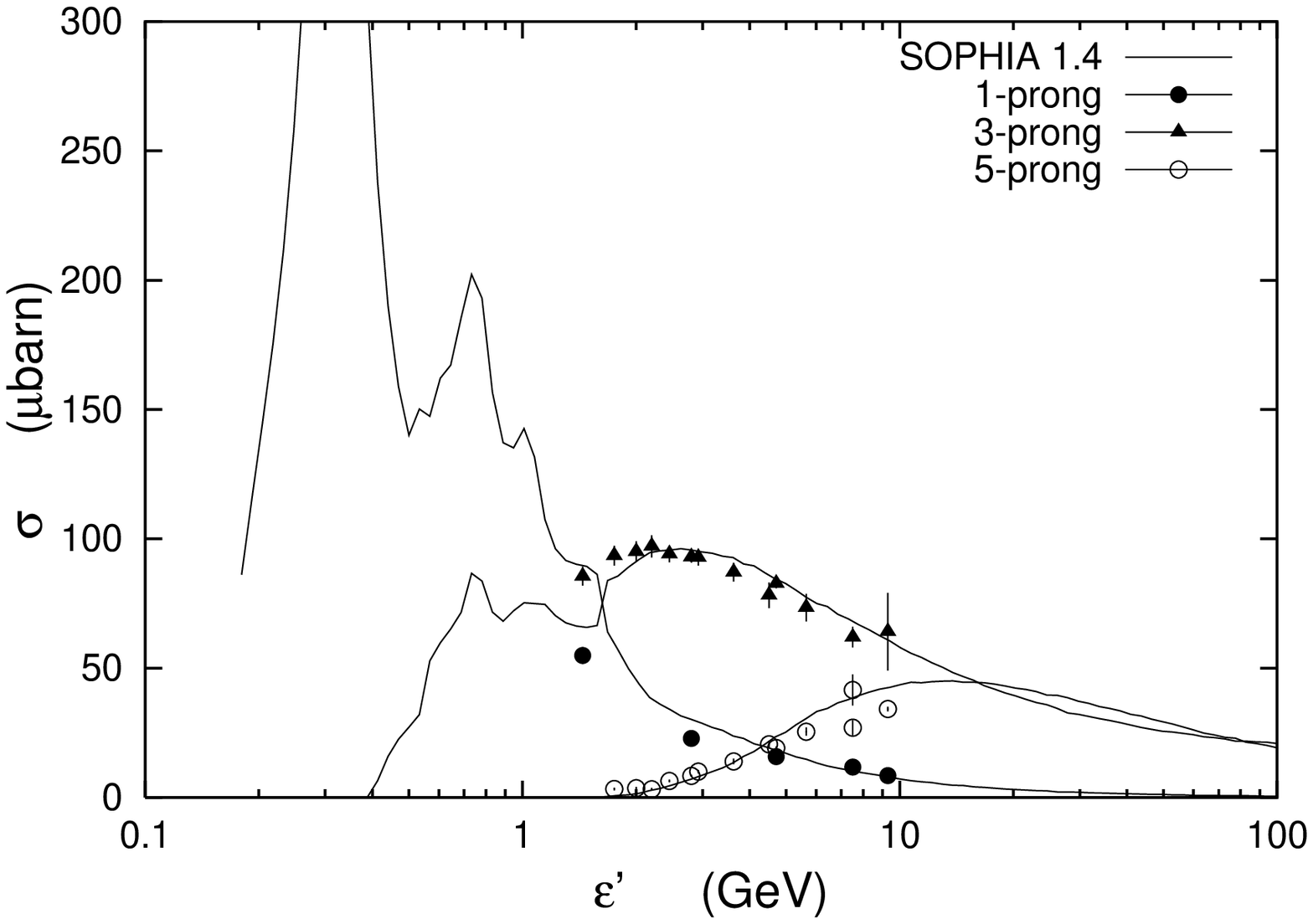,width=13cm,height=10cm}}
\caption[]{
\label{pion-spec-2}
Cross section for n-prongs (n=1,2,3).
Data are from \cite{ball72,alex65,bing73,stru76,ball69b}.}
\end{figure}

\begin{figure}[!htb]
{\psfig{figure=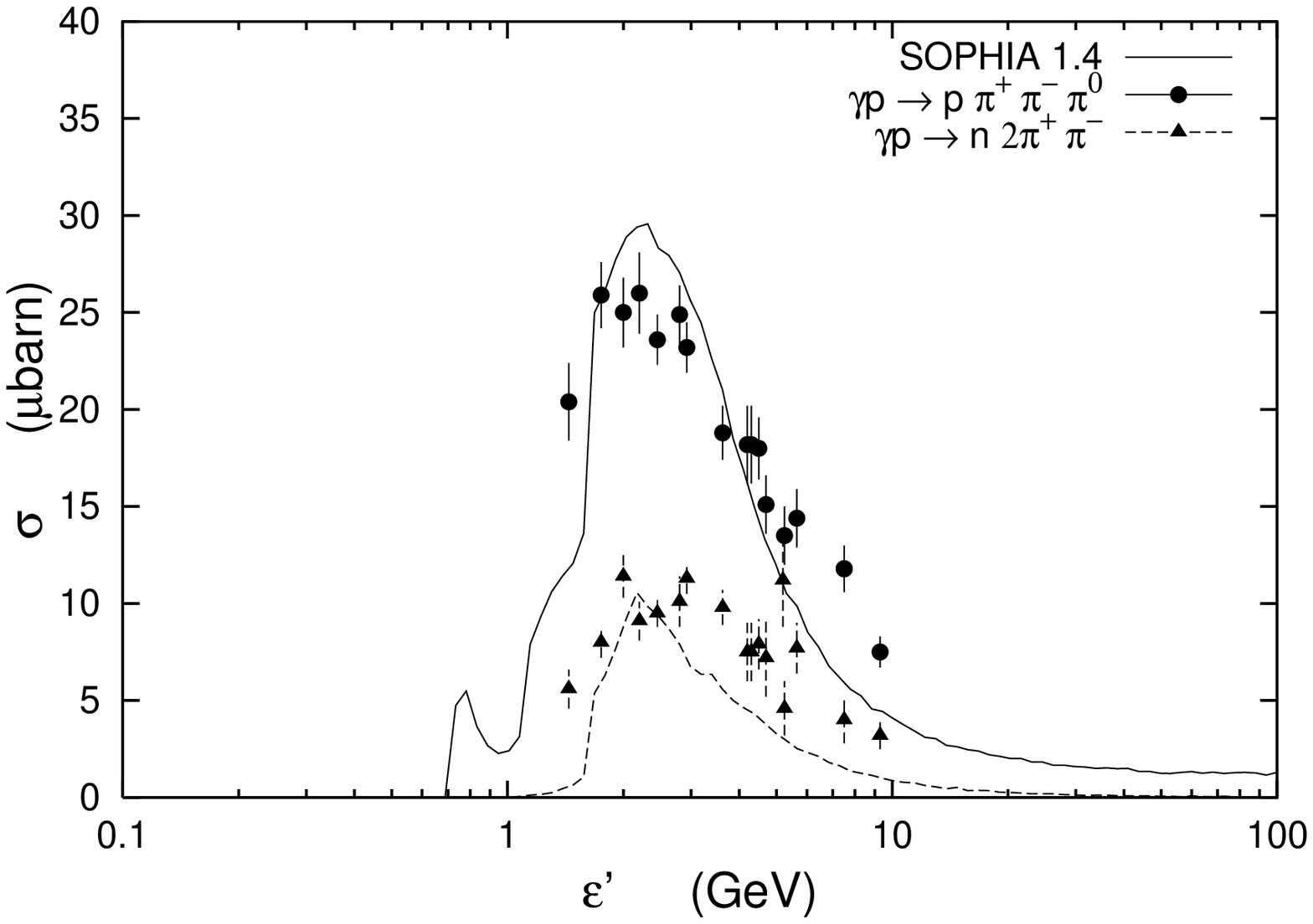,width=14cm,height=7.5cm}}
{\psfig{figure=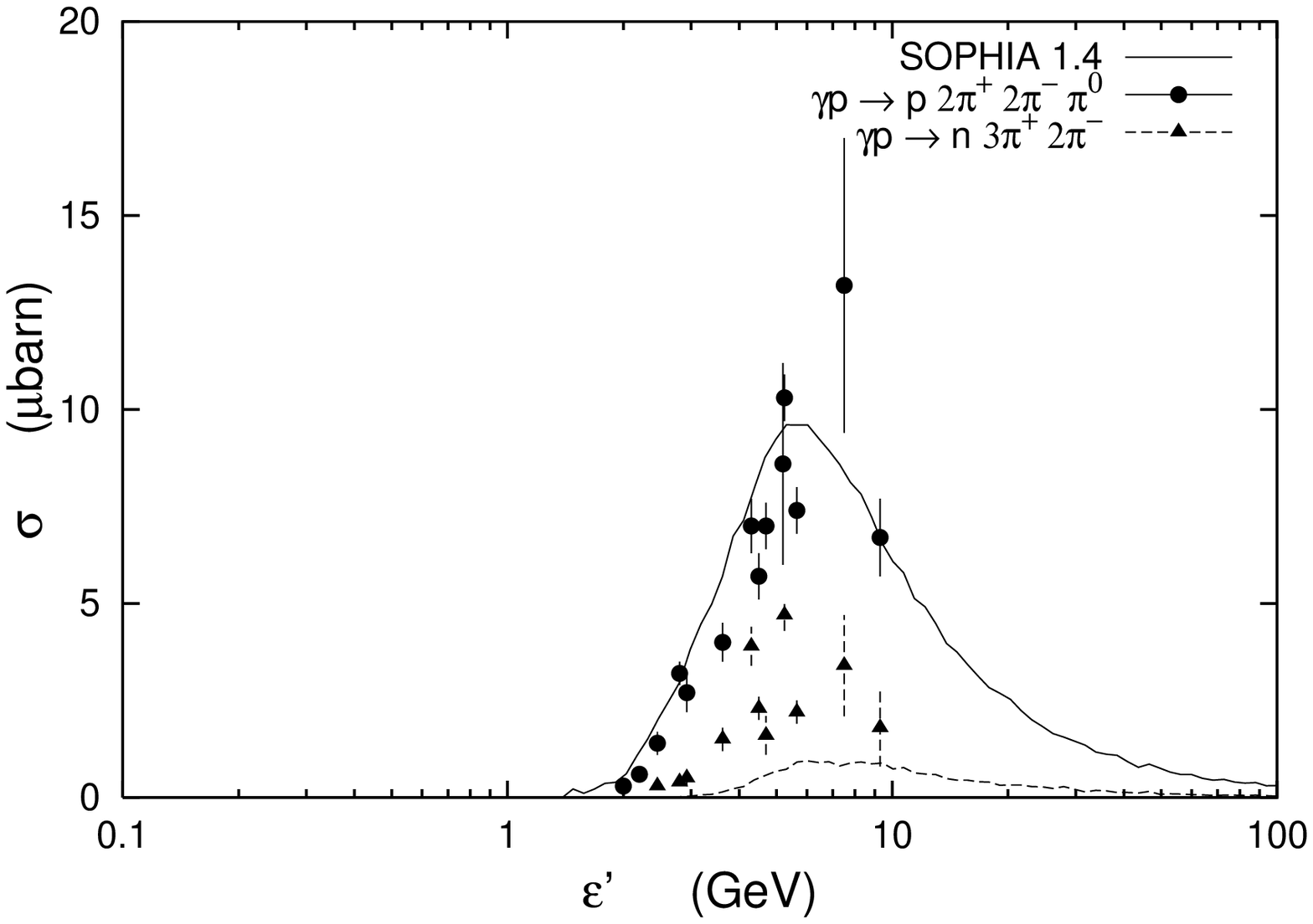,width=14cm,height=7.5cm}}
\caption[]{
\label{pion-spec-3}
Total cross section of $\gamma p\longrightarrow \pi^+\pi^-\pi^0 p$ 
in comparison with
$\gamma p\longrightarrow 2\pi^+\pi^- n$
and $\gamma p\longrightarrow 3\pi^+3\pi^- n$ in comparison with
$\gamma p\longrightarrow 2\pi^+2\pi^-\pi^0 p$. 
Data are from \cite{ball72,stru76,ball68,ball69a,bing73,carb76,haus,eise72,%
alex65}.}
\end{figure}

Figs.~\ref{pion-spec-0} to \ref{pion-spec-3} 
compare the output of SOPHIA with the data 
on specific final states as a function of the interaction energy.
 Such comparisons are important for models that aim
 to represent correctly photohadronic interactions over a wide
 energy range. 
 
Fig.~\ref{pion-spec-0} compares the total cross sections for
$\pi^0 p$ and $\pi^+ n$ production with experimental data.
The major contributions come from the
$\Delta$(1232) resonance and the direct channel
together with minor contributions from other
resonances. The agreement with data in the threshold region
is of great importance for many astrophysical applications
where this is the dominating energy range in the case of steep
proton and ambient photon spectra.

Fig.~\ref{pion-spec-1} compares the calculated and measured cross sections
for final states involving charged and neutral pions.
The general agreement of the model results 
with data is quite good, although there are some
energy ranges that show minor deviations.
It is important to note that it is difficult to fit
exactly the experimental data without any detailed knowledge of the
experimental setups and acceptance constraints, especially in cases like 
$\gamma p \rightarrow \pi^+ \pi^- p$ + neutrals, where the final state  
is not well defined.

The number of inelastic $\gamma p$ events with 1, 3, and 5 charged 
particles in the final state are shown in Fig.~\ref{pion-spec-2}.
This comparison
is very sensitive to the description of multipion production processes
as well as to the smooth transition between different particle
production processes.
It shows that the
 different channels are reasonably well modeled in SOPHIA.  

Fig.~\ref{pion-spec-3} shows reaction cross sections for final states
with equal numbers of pions, 
but having a different isospin of the produced nucleon,
together with fixed-target data. 
This comparison is important for the correct simulation of the ratio of
the proton to neutron numbers produced in $\gamma p$ collisions.
Again the model results agree well with data.

\subsection{Inclusive distributions}

 The rapidity distribution tests the kinematic of our simulations. 
The rapidity of a final state particle with energy $E$ is defined as
\begin{equation}
y = \frac{1}{2} \ln \left(\frac{E+p_\parallel}{E-p_\parallel}\right)
= \ln \left(\frac{E+p_\parallel}{m_\perp}\right),
\label{rapidity}
\end{equation}
where $p_\parallel$ is its momentum component along the direction of the
incoming particle. The transverse mass $m_\perp$ follows from
$m_\perp^2 = E^2-p_\parallel^2$.
Rapidity is additive under Lorentz transformations, which keeps the
rapidity distribution invariant under such transformations.

\begin{figure}[!htb]
{\psfig{figure=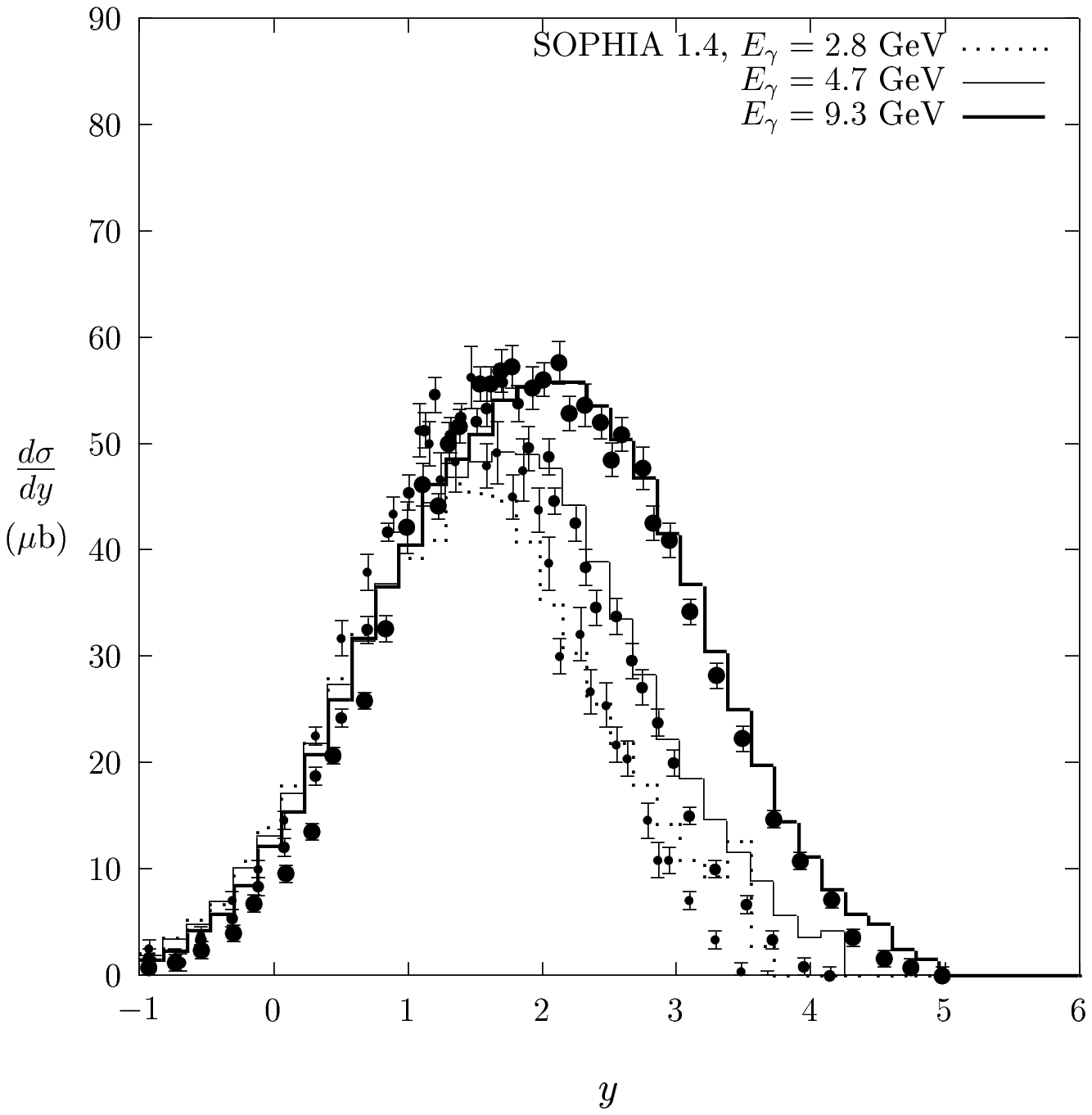,width=14cm,height=10cm}}
\caption[]{
\label{rapidity-spec}
Rapidity distribution for 
$\gamma p\longrightarrow \pi^-+$anything at beam energies 
$\epsilon^\prime = 2.8$GeV (dotted line), 
4.7 GeV (solid line) and 9.3 GeV (thick solid line).
Data are from Moffeit et al.~\cite{moff72}.}
\end{figure}

Moffeit et al.~\cite{moff72} have measured the rapidity
distribution for the interaction 
$\gamma p\longrightarrow \pi^- +$ anything at three
different beam energies $\epsilon^{\prime}$ = 2.8, 4.7 and 9.3 GeV.
In Fig.~\ref{rapidity-spec} we compare the calculated rapidity distribution
to data. The agreement in the width and the height of the distributions 
is good.

\subsection{Multiplicities}

\begin{figure}[!htb]
\begin{minipage}[t]{7.4cm}
{\psfig{figure=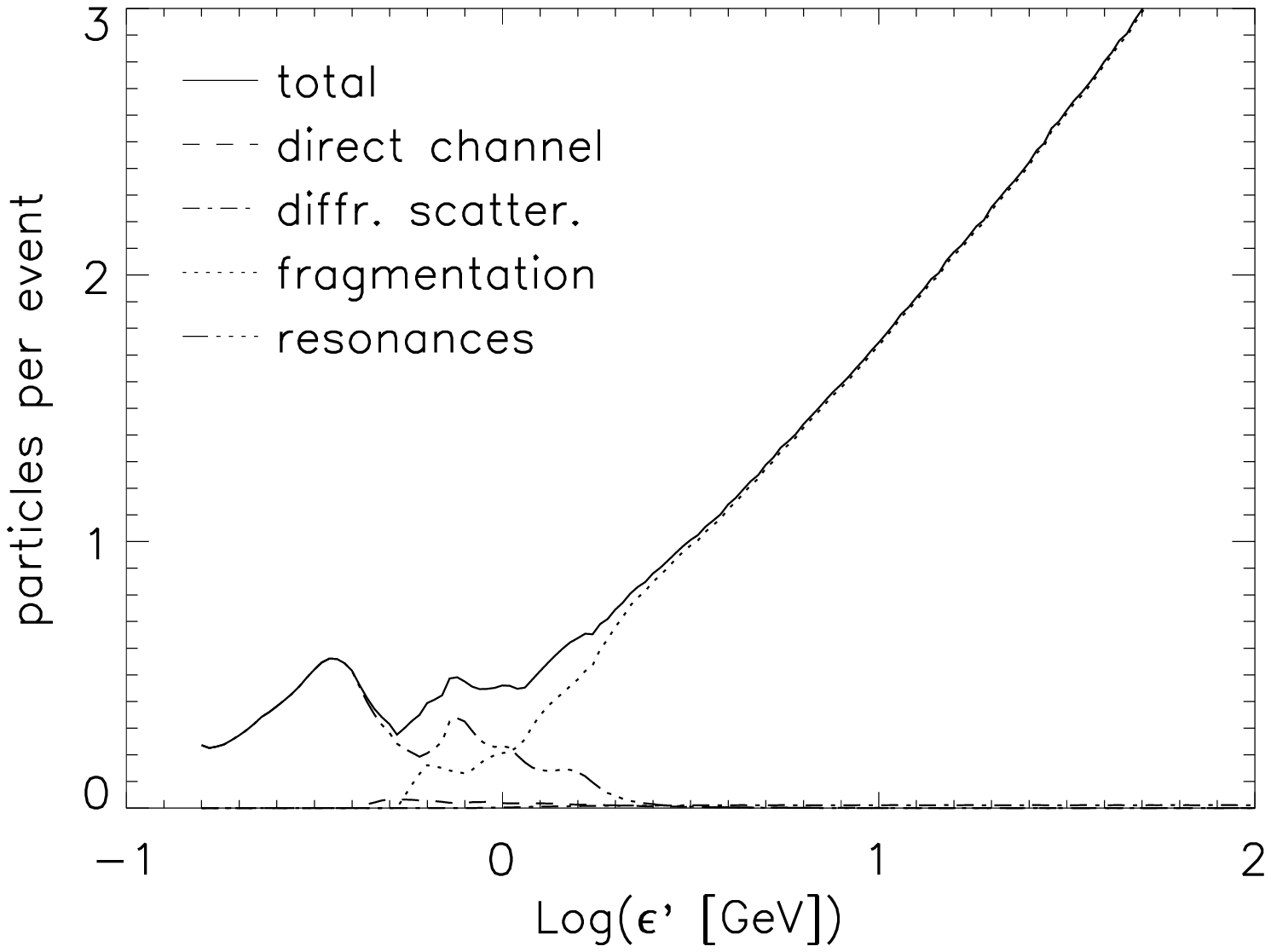,height=7cm,width=7.3cm}}
\end{minipage}
\hfill
\begin{minipage}[t]{7.4cm}
\vspace*{-7.cm}
{\psfig{figure=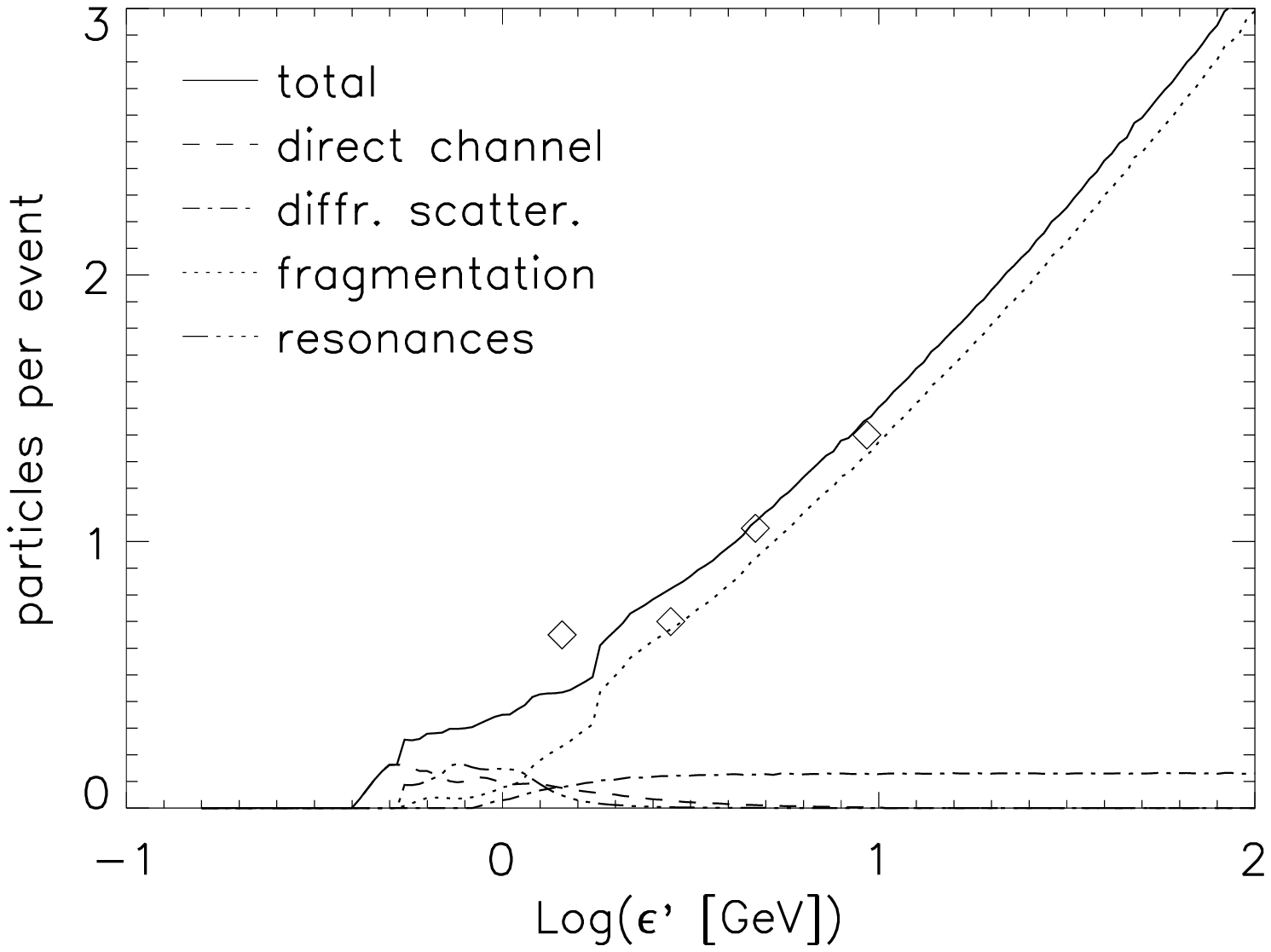,height=7cm,width=7.3cm}}
\end{minipage}
\caption[]{
\label{pion-mul-1}
Left figure: $\pi^0$-multiplicity for $\gamma p$-interactions.
Right figure: $\pi^-$-multiplicity for $\gamma p$-interactions. 
Data (diamonds) are from Moffeit et al.~\cite{moff72}.}
\end{figure}

\begin{figure}[!htb]
\begin{minipage}[t]{7.4cm}
{\psfig{figure=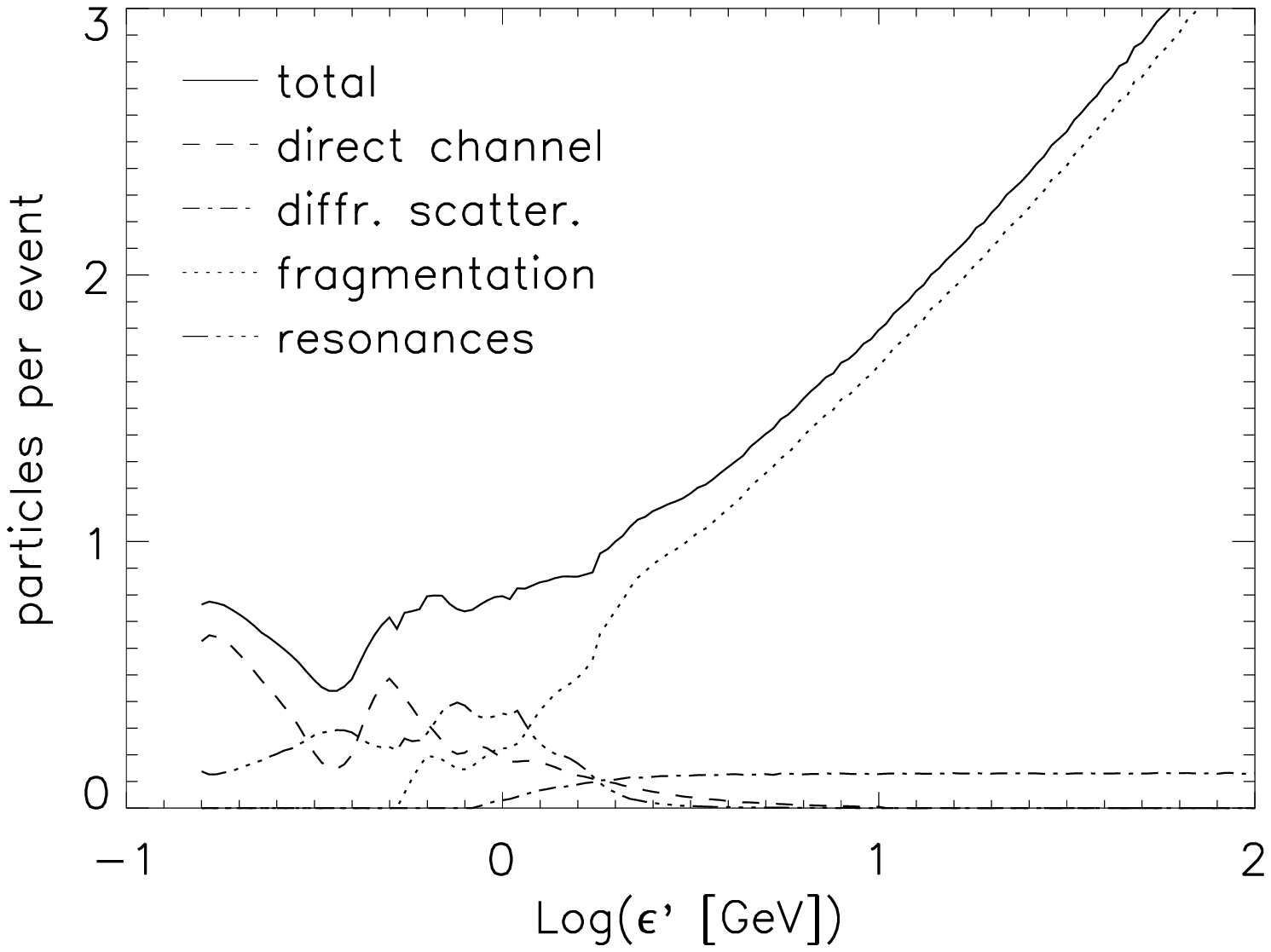,height=7cm,width=7.3cm}}
\end{minipage}
\hfill
\begin{minipage}[t]{7.4cm}
\vspace*{-7.cm}
{\psfig{figure=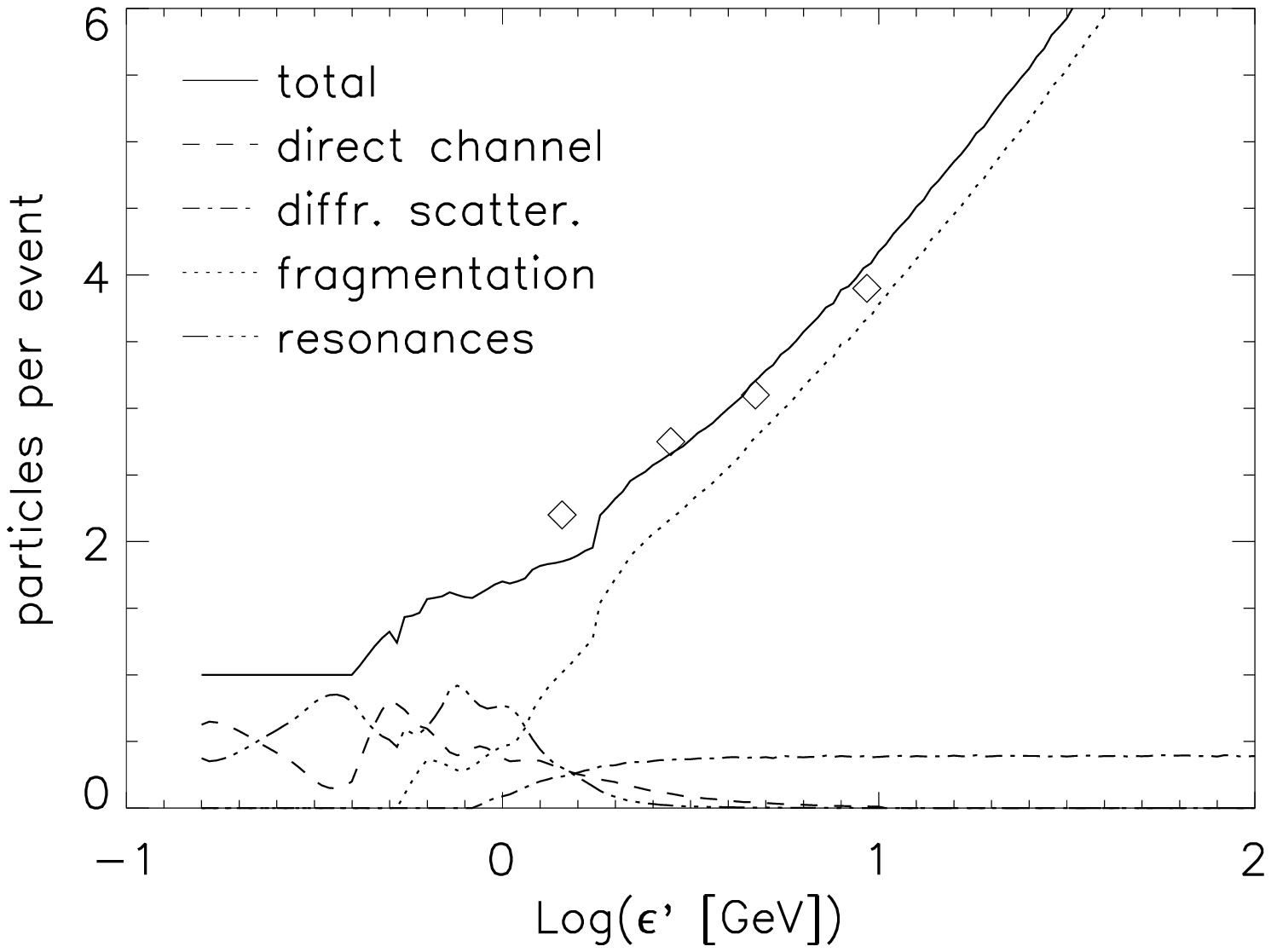,height=7cm,width=7.3cm}}
\end{minipage}
\caption[]{
\label{pion-mul-2}
Left figure: $\pi^+$-multiplicity for $\gamma p$-interactions.
Right figure: Charged particle multiplicity for 
$\gamma p$-interactions. Data (diamonds) 
are from Moffeit et al.~\cite{moff72}.}
\end{figure}

In astrophysical environments the production of neutrinos 
results mainly from the decay of charged secondary pions 
($\pi^+\rightarrow e^+\nu_{\mu}\bar\nu_{\mu}\nu_e$, 
$\pi^-\rightarrow e^-\nu_{\mu}\bar\nu_{\mu}\bar\nu_e$).
Most of the gamma rays are produced via neutral pion 
decay ($\pi^0\rightarrow\gamma\gamma$) 
and synchrotron/Compton emission from emanating 
relativistic $e^-/e^+$.
Pion multiplicities (see Figs.~\ref{pion-mul-1}, \ref{pion-mul-2}) 
are therefore instructive to understand the energy dissipation of
the initial energy among the $\nu$- and $\gamma$-component.
For non-astrophysical applications the charged pion multiplicity is a basic
interaction parameter that presents a cumulative measure of many
interaction channels. 

In the resonance region
the maximum of the neutral pion multiplicity is reached at 
the $\Delta(1232)$-resonance (see Fig.~\ref{pion-mul-1}).
At threshold neutral pion production is strongly suppressed 
in favour of $\pi^+$-production (see Fig.~\ref{pion-mul-2})
due to the dominance of the direct channel.
Multiplicities are approximately growing as $\sim s^{1/4}$ in the
multipion production region at low energies.
The obtained multiplicity distributions from our simulations are 
in agreement with the data from Moffeit et al.~\cite{moff72}.

\begin{figure}[!htb]
\centering
{\psfig{figure=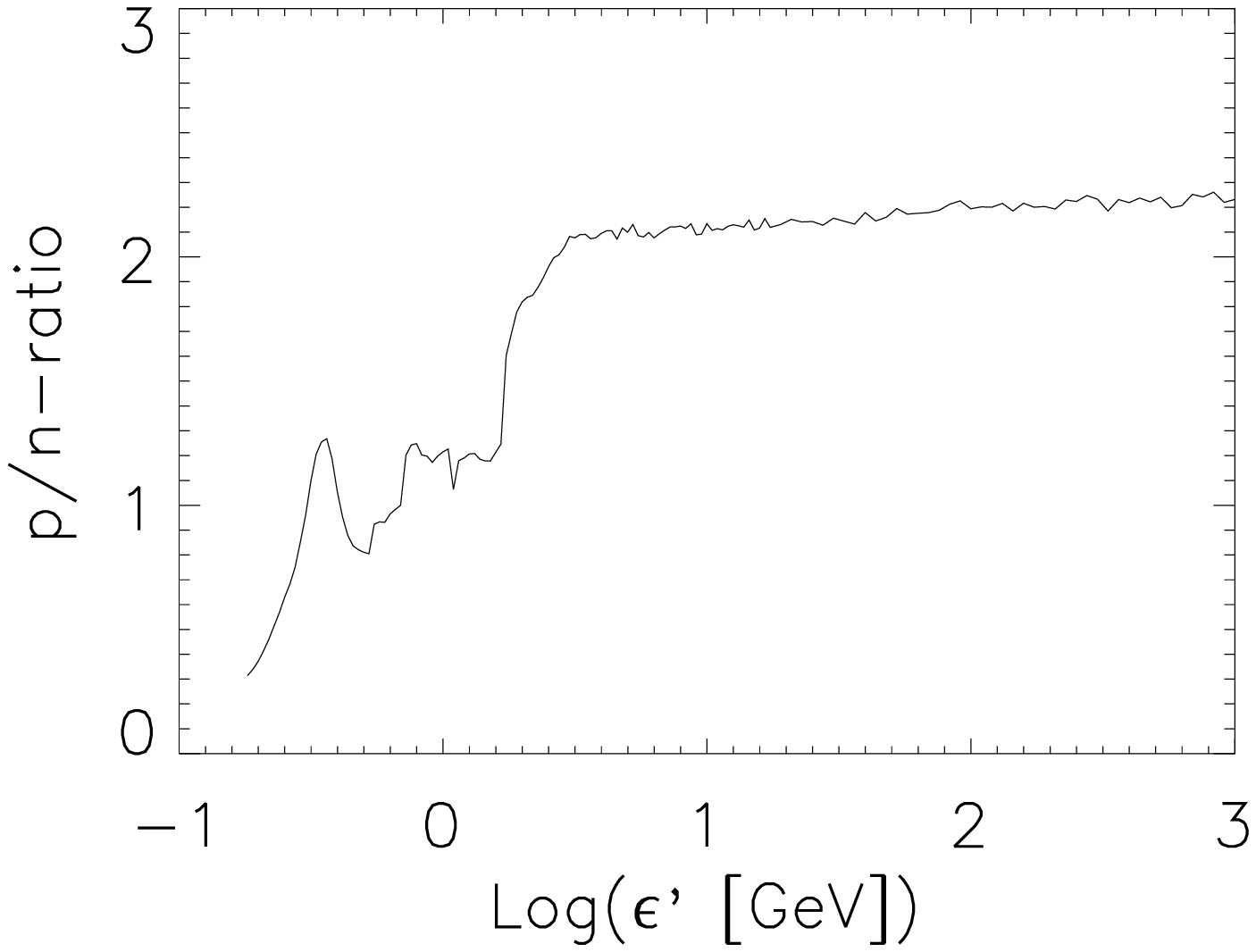,width=10cm,height=8cm}}
\caption[]{
\label{proton-to-neutron}
Proton-to-neutron ratio for 
$\gamma p$-interactions as simulated by SOPHIA. At high energy
this ratio is $\sim 2.2$. 
}
\end{figure}
By counting the number of protons and neutrons produced in $\gamma p$
interactions, one can define a 
proton-to-neutron ratio.
The SOPHIA prediction on the energy dependence of this ratio is shown 
in Fig.~\ref{proton-to-neutron}.
The $p/n$ ratio reaches 
 $\approx 2.2$ at high energy 
which can be contrasted to an experimentally estimated value of 
about $3.8$ found by Meyer \cite{mey73}.
The ratio derived in \cite{mey73} is essentially based on
the same data as those our model is compared to. Meyer estimated the
 experimentally unknown cross sections by isospin symmetry
 arguments. In our case these cross sections are predicted by the Monte
 Carlo simulation. We conclude from this that the unmeasured cross
 sections are important for the $p/n$ ratio, and that the difference between 
the values 2.2 and 3.8
reflect the uncertainty due to the limited experimental data available.

\section{Conclusions}

A newly developed Monte Carlo event generator SOPHIA has been presented. 
It simulates the interactions of nucleons
with photons over a wide range in energy.
The simulation of the final state includes
 all interaction processes which are relevant to astrophysical applications.
SOPHIA contains tools,
 such as for sampling the photon energy from different ambient soft
 photon spectra, and the nucleon--photon interaction angle, that are
 needed for such applications. As an event generator, SOPHIA uses
 all available information on the interaction cross section, the final state
 particle composition, and kinematics of the interaction processes as
 provided by particle physics. 
Comparison with the available accelerator data shows that SOPHIA
provides a good description of our current knowledge
of photon--nucleon interactions. 

\vspace*{1cm}

\begin{minipage}[t]{15cm}
\centerline{\bf{Acknowledgements}}
 
The work of AM and RJP is supported by the Australian Research Council.
RE and TS acknowledge the support by the U.S. Department of Energy under 
Grant Number DE FG02 01 ER 40626. TS is also supported in part by
the NASA grant NAG5--7009. The contribution of JPR was supported by NASA
NAG5--2857 and by the EU-TMR network ``Astro-Plasma-Physics'' under the
contract number ERBFMRX-CT98-0168.
\end{minipage}

\clearpage



\begin{appendix}

\section{Definition of functions\label{app:functions}}

The functions $\Pl(\eps',\eps'_{\rm th},\eps'_{\rm max},\alpha)$ and
$\Qf(\eps',\eps'_{\rm th},w)$ are defined by 
\begin{equation}
\Pl(\eps',\eps'_{\rm th},\eps'_{\rm max},\alpha) = (\frac{\eps'-\eps'_{\rm
 th}}{\eps'_{\rm max}-\eps'_{\rm th}})^{A-\alpha} (\frac{\eps'}{\eps'_{\rm
 max}})^{-A}
\label{Pl}
\end{equation}
for $\eps'>\eps'_{\rm th}$, and $\Pl(\eps',\eps'_{\rm th},\eps'_{\rm
 max},\alpha) = 0$ otherwise, $A=\alpha \eps'_{\rm max}/{\eps'_{\rm th}}$, and
\begin{equation}
\Qf(\eps',\eps'_{\rm th},w) = (\eps'-\eps'_{\rm th})/w
\label{Qf}
\end{equation}
for $\eps'_{\rm th}<\eps'<w+\eps'_{\rm th}$, $\Qf(\eps',\eps'_{\rm th},w) =
0$ for $\eps'\leq \eps'_{\rm th}$, and $\Qf(\eps',\eps'_{\rm th},w) = 1$ for
$\eps'\leq w+\eps'_{\rm th}$.

\clearpage

\section{Resonance branching ratios\label{app:branches}}
\begin{small}
\begin{table}[!htb]
\caption[]{
\label{tab:resbranch}
Decay channels and branching ratios $b_c$ for resonance decays, following the
scheme described in the text.  The second row of each table gives the
fractional contribution of the decay probability for several given energy
ranges. The average branching ratio for each channel is given in the last
column of each table. They are generally consistent with the average values
quoted in the \RPP~\cite{Caso98a}, except for the $N\rho$ decay of the
$N(1520)$, where $b_{N\rho} \le 6\%$ due to the low fractional contribution
of this energy range whereas $15\%{-}25\%$ are quoted in the \RPP.
}
\begin{center}
\renewcommand{\arraystretch}{1.2}
\begin{tabular}{c|c|r} \hline
$\Delta$(1232) 	& all x 	& total \\\hline
     fraction   & 100\%		& \\\hline 
N$\pi$    	& 100\%  	& 100\%
\end{tabular}

\vspace*{3mm}

\begin{tabular}{c|cc|r} \hline
N(1440) 	& x$<$0.54 	& x$\ge$0.54 	& total \\\hline
     fraction   & 34\%		& 66\%		& \\\hline 
N$\pi$    	& 100\%  	& 50\%		& 67\%\\
$\Delta\pi$  	& -		& 50\%		& 33\%\\
\end{tabular}

\vspace*{3mm}

\begin{tabular}{c|ccc|r} \hline
N(1520) 	& x$<$0.54 	& 0.54$\le$x$<$1.09 & x$\ge$1.09 & 
total  \\\hline
     fraction   & 9\%		& 85\%		&  6\%		& \\\hline 
N$\pi$    	& 100\%  	& 50\%		&   0\%		& 52\%\\
$\Delta\pi$  	& -		& 50\%		&   0\%		& 42\%\\
N$\rho$		& -		& -		& 100\%		& 6\%\\
\end{tabular}

\vspace*{3mm}

\begin{tabular}{c|cc|r} \hline
N(1535) 	& x$<$0.71 	& x$\ge$0.71 	& total \\\hline
    fraction    & 31\%		& 69\%		& \\\hline 
N$\pi$    	& 100\%  	& 25\%		& 45\%\\
N$\eta$	  	& -		& 75\%		& 55\%\\
\end{tabular}

\vspace*{3mm}

\begin{tabular}{c|ccc|r} \hline
N(1650) 	& x$<$0.54 	& 0.54$\le$x$<$0.91 & x$\ge$0.91
& total \\\hline
fraction        & 5\%		& 26\%		&  69\%		& \\\hline 
N$\pi$    	& 100\%  	& 85\%		&  70\%		& 75\%\\
$\Delta\pi$  	& -		& 15\%		&  15\%		& 14\%\\
$\Lambda$K	& -		& -		&  15\%		& 11\%\\
\end{tabular}

\end{center}

\end{table}

\clearpage

\begin{table}[!htb]

\begin{center}

\renewcommand{\arraystretch}{1.2}
\begin{tabular}{c|cc|r} \hline
N(1675) 	& x$<$0.54 	& x$\ge$0.54 	& total \\\hline
fraction        & 5\%		& 95\%		& \\\hline 
N$\pi$    	& 100\%  	& 40\%		& 42\%\\
$\Delta\pi$  	& -		& 60\%		& 57\%\\
\end{tabular}

\vspace*{3mm}

\begin{tabular}{c|ccc|r}\hline
N(1680) 	& x$<$0.54 	& 0.54$\le$x$<$1.09 & x$\ge$1.09
& total \\\hline
    fraction    & 4\%		& 64\%		&  32\%		& \\\hline 
N$\pi$    	& 100\%  	& 65\%		&  55\%		& 64\%\\
$\Delta\pi$  	& -		& 35\%		&  0\%		& 22\%\\
N$\rho$		& -		& -		&  45\%		& 14\%\\
\end{tabular}

\vspace*{3mm}

\begin{tabular}{c|ccc|r}\hline
$\Delta$(1700) 	& x$<$0.54 	& 0.54$\le$x$<$1.09 & x$\ge$1.09
& total \\\hline
    fraction    & 9\%		&  52\%		&  39\%		& \\\hline 
N$\pi$    	& 100\%  	&   0\%		&   0\%		& 14\%\\
$\Delta\pi$  	& -		& 100\%		&  20\%		& 55\%\\
N$\rho$		& -		& -		&  80\%		& 31\%\\
\end{tabular}

\vspace*{3mm}

\begin{tabular}{c|ccc|r}\hline
$\Delta$(1905) 	& x$<$1.09 	& 0.54$\le$x$<$1.09 & x$\ge$1.09		
& total\\\hline
fraction        & 6\%		& 21\%		&  73\%		& \\\hline 
N$\pi$    	& 100\%  	& 40\%		&   0\%		& 14\%\\
$\Delta\pi$  	& -		& 60\%		&   0\%		& 13\%\\
N$\rho$	  	& -		& - 		& 100\%		& 73\%\\
\end{tabular}

\vspace*{3mm}

\begin{tabular}{c|ccc|r}\hline
$\Delta$(1950) 	& x$<$0.54 	& 0.54$\le$x$<$1.09 & x$\ge$1.09 & 
total\\\hline
fraction   & 4\%		& 15\%		&  81\%		& \\\hline 
N$\pi$    	& 100\%  	& 60\%		&  30\%		& 37\%\\
$\Delta\pi$  	& -		& 40\%		&  40\%		& 39\%\\
N$\rho$		& -		& -		&  30\%		& 24\%\\
\end{tabular}

\end{center}

\end{table}

\end{small}

\clearpage
  
\section{Compilation of routines/functions}

\begin{description}
\item[function {\tt BREITWIGNER(SIGMA\_0, GAMMA, DMM,
EPS\_PRIME)}:]\hspace*{1cm}\\
calculates cross section (in $\mu$barn) of a resonance with width {\tt GAMMA}
 (in GeV),
mass {\tt DMM} (in GeV), 
maximum cross section {\tt SIGMA\_0} (in $\mu$barn) and NRF energy
of the photon (in GeV) according to the Breit-Wigner formula
\item[subroutine {\tt CROSSDIR(EPS\_PRIME)}:]\hspace*{1cm}\\
    collection of functions ({\tt SINGLEBACK, TWOBACK}) which calculates the
    cross section (at the NRF energy {\tt EPS\_PRIME} 
    of the incident photon) of the direct
    channel (not isospin-corrected)
\item[function {\tt CROSSECTION(EPS\_PRIME,NDIR)}:]\hspace*{1cm}\\
    computes cross section (in $\mu$barn) for $N\gamma$-interaction at a given
    energy {\tt EPS\_PRIME} (=photon energy in GeV in proton's rest frame);
    depending on the control variable {\tt NDIR} it returns the total
    cross section ({\tt NDIR=3}) or only a certain part of the cross section
    ({\tt NDIR=1}: total resonance cross section, {\tt NDIR=4}: 
direct channel cross section
    {\tt NDIR=5}: fragmentation cross section, {\tt NDIR=11-19}: 
individual resonance cross sections)
\item[subroutine {\tt DEC\_INTER3(EPS\_PRIME,IMODE)}:]\hspace*{1cm}\\
    returns reaction mode: decay of resonance ({\tt IMODE=6}), direct pion
 production ({\tt IMODE=2}
    for $N\pi$ final states, {\tt IMODE=3} for $\Delta\pi$ final states),
 fragmentation in resonance region
    ({\tt IMODE=5}), diffractive scattering
    ({\tt IMODE=1} for $N\rho$ final states, 
     {\tt IMODE=4} for $N\omega$ final states)
    and multipion production/fragmentation ({\tt IMODE=0}) at a given energy
 {\tt EPS\_PRIME} (in GeV)
\item[subroutine {\tt
DEC\_PROC2(EPS\_PRIME,IPROC,IRANGE,IRES,L0)}:]\hspace*{1cm}\\
    returns in {\tt IPROC} the decay mode for a given resonance {\tt IRES} at 
    energy {\tt EPS\_PRIME} (in GeV) for incident nucleon with code 
    number {\tt L0};
    {\tt IRANGE} is the number of energy intervals corresponding to 
    the (energy dependent)
    branching ratios of a specific resonance
\item[subroutine {\tt DEC\_RES2(EPS\_PRIME,IRES,IRESMAX,L0)}:]\hspace*{1cm}\\
    returns sampled resonance number {\tt IRES} ({\tt IRES = 1 \ldots IRESMAX}) 
    at energy
    {\tt EPS\_PRIME} (in GeV) for incident nucleon with code number {\tt L0}
\item[subroutine {\tt DECSIB}:]\hspace*{1cm}\\
    decay of unstable particles
\item[function {\tt FUNCTS(S)}:]\hspace*{1cm}\\
    calculates distribution of the squared CMF energy {\tt S} (in GeV$^2$)
\item[subroutine {\tt GAMMA\_H(E0,L0,IMODE)}:]\hspace*{1cm}\\ 
    interface routine for hadron-$\gamma$ collisions in CM frame;\\
    {\tt E0} is CMF energy of the $N\gamma$-system; first particle ($\gamma$ or
    hadron $N$) goes in $+z$-direction; final state consists of protons, 
    neutrons, gamma, leptons, neutrinos
\item[routine {\tt INITIAL(L0)}:]\hspace*{1cm}\\
     initialization routine for parameter settings;\\ 
     to be called at before calling {\tt EVENTGEN}
\item[routine {\tt LUND\_FRAG(SQS,NP)}:]\hspace*{1cm}\\
     interface to {\sc Jetset} fragmentation of system with 
     CMF energy {\tt SQS} (in GeV);\\
     {\tt NP} = number of secondaries produced
\item[function {\tt PHOTD(EPS,TBB)}:]\hspace*{1cm}\\
     returns photon density (in photons/cm$^3$/eV) at energy 
     {\tt EPS} (in eV) of 
     blackbody radiation of temperature {\tt TBB} (in K)
\item[subroutine {\tt PROBANGLE(IRES,L0,ANGLESCAT)}:]\hspace*{1cm}\\
    probability distribution for scattering angle of given resonance 
    {\tt IRES} and incident nucleon with code number {\tt L0} ;\\
    {\tt ANGLESCAT} is the cosine of scattering angle in CMF frame
\item[function {\tt PROB\_EPSKT(EPS)}:]\hspace*{1cm}\\
     calculates probability distribution for thermal photon field 
     with temperature {\tt TBB} (in K) at energy {\tt EPS} (in eV)
\item[function {\tt PROB\_EPSPL(EPS)}:]\hspace*{1cm}\\
     calculates probability distribution at energy {\tt EPS} (in eV) 
     for power law 
     radiation n $\sim$ {\tt EPS}$^{-{\tt ALPHA}}$ between the 
     limits {\tt EPSM1 \ldots EPSM2} 
     (in eV)
\item[subroutine {\tt PROC\_TWOPART(LA,LB,AMD,LRES,PRES,COSTHETA,NBAD)}:]
    carries out 2-particle decay in CM frame of system with mass {\tt AMD} 
    (in GeV) into 
    particles with code numbers {\tt LA,LB} (stored in array {\tt
    LRES()}) wheras
    {\tt COSTHETA} is the cosine of the $N\gamma$-CM frame 
    scattering angle ; returns 5-momenta {\tt PRES()} of the two particles;\\
    {\tt NBAD=1} for kinematically not possible decays, 
    otherwise {\tt NBAD=0}
\item[subroutine {\tt
RES\_DECAY3(IRES,IPROC,IRANGE,S,L0,NBAD)}:]\hspace*{1cm}\\
    determines decay products for the decay of a given resonance 
    {\tt IRES} by a
    given decay process {\tt IPROC} for a incident nucleon with code number 
    {\tt L0},
    and carries out two-particle decay of the
    resonance with squared CMF energy {\tt S};
    code numbers and (CMF) 5-momenta of produced particles are stored in arrays
    {\tt LLIST} and {\tt P}, respectively, in a common block;\\
    {\tt NBAD=1} for kinematically not possible decays; otherwise {\tt NBAD=0}
\item[subroutine {\tt SAMPLE\_EPS(EPS,EPSMIN,EPSMAX)}:]\hspace*{1cm}\\
    samples incident photon energy; if the blackbody temperature is set to 
    {\tt TBB=0} the
    photon energy is sampled from a power law distribution 
    with index {\tt ALPHA} between given photon 
    energies {\tt EPSMIN, EPSMAX} (in eV); 
    otherwise it is sampled from a distribution for a blackbody radiation with
    temperature {\tt TBB} (in K)
\item[subroutine {\tt SAMPLE\_S(S,EPS)}:]\hspace*{1cm}\\
    samples the total CM frame energy {\tt S} (in GeV$^2$) for a given photon 
    with energy {\tt EPS} (in GeV) and proton with energy {\tt E0} (in GeV)
\item[subroutine {\tt SCATANGLE(ANGLESCAT,IRES,IPROC,INC)}:]\hspace*{1cm}\\
    samples the cosine of the scattering angle {\tt ANGLESCAT} for a given
    resonance {\tt IRES} and incident nucleon {\tt INC}
\end{description}

\end{appendix}

\end{document}